\documentclass[sigconf,nonacm]{acmart}
\pdfoutput=1
\renewcommand\footnotetextcopyrightpermission[1]{} 
\usepackage{amssymb,amsmath,epsfig,graphics,enumerate,float,subcaption,wrapfig,balance,xspace,url}
\usepackage{color}
\usepackage{booktabs}
\usepackage{graphicx,pifont}

\AtBeginDocument{%
  \providecommand\BibTeX{{%
    \normalfont B\kern-0.5em{\scshape i\kern-0.25em b}\kern-0.8em\TeX}}}

\setcopyright{none}
\copyrightyear{2020}

\hypersetup{draft}
\begin{document}

\title{Road Grade Estimation Using Crowd-Sourced Smartphone Data}


\author{Abhishek Gupta}
\affiliation{\institution{SUNY Buffalo, NY, USA}}
\email{agupta33@buffalo.edu}

\author{Shaohan Hu}
\affiliation{\institution{IBM Research, Yorktown Heights, NY, USA}}
\email{shaohan.hu@ibm.com}

\author{Weida Zhong}
\affiliation{\institution{SUNY Buffalo, NY, USA}}
\email{weidazho@buffalo.edu }

\author{Adel Sadek}
\affiliation{\institution{SUNY Buffalo, NY, USA}}
\email{asadek@buffalo.edu}

\author{Lu Su}
\affiliation{\institution{SUNY Buffalo, NY, USA}}
\email{lusu@buffalo.edu}

\author{Chunming Qiao}
\affiliation{\institution{SUNY Buffalo, NY, USA}}
\email{qiao@buffalo.edu}

\renewcommand{\shortauthors}{Gupta et al.}

\begin{abstract}
 Estimates of road grade/slope can add another dimension of information to existing 2D digital road maps. Integration of road grade information will widen the scope of digital map's applications, which is primarily used for navigation, by enabling driving safety and efficiency applications such as Advanced Driver Assistance Systems (ADAS), eco-driving, etc. The huge scale and dynamic nature of road networks make sensing road grade a challenging task. Traditional methods oftentimes suffer from limited scalability and update frequency, as well as poor sensing accuracy. To overcome these problems, we propose a cost-effective and scalable road grade estimation framework using sensor data from smartphones. Based on our understanding of the error characteristics of smartphone sensors, we intelligently combine data from accelerometer, gyroscope and vehicle speed data from OBD-II/smartphone's GPS to estimate road grade. To improve accuracy and robustness of the system, the estimations of road grade from multiple sources/vehicles are crowd-sourced to compensate for the effects of varying quality of sensor data from different sources. Extensive experimental evaluation on a test route of ~9km demonstrates the superior performance of our proposed method, achieving $5\times$ improvement on road grade estimation accuracy over baselines, with 90\% of errors below 0.3$^\circ$.

\end{abstract}

\maketitle

\section{Introduction}
Grade is the measure of the road's steepness as it rises and falls along its route. The key objective in designing grade profiles of road is to optimize its efficiency and safety. As the automobile sector is gradually moving towards complete autonomy, there is an ever-increasing requirement for scalable and cost efficient solutions to collect rich and accurate data about road networks. Especially, gathering and estimating accurate road grade information will enrich existing 2D digital maps, thus widening the scope of its applicability. For example, road grade information can be used to determine fuel efficient routes between a source and destination~\cite{liu2010fuel}. Other applications include eco-driving~\cite{kang2015ecodrive}, Advance Driver Assistance systems (ADAS)~\cite{adas}, 3D-Object detection for autonomous vehicles~\cite{yang2018hdnet} and HD-Map creation~\cite{hdmaps} etc.

The task of estimating road grade is challenging due to the sheer scale of the road networks, as well as their dynamic nature stemmed from the construction of new roads and maintenance of the existing ones. According to US Department of Transportation (USDOT), between 2000 and 2016, the U.S. built an average of 30,427 lane miles of roadway per year~\cite{usdot}. In addition to new roads, existing roads are maintained/modified for accident vulnerability prevention, traffic flow enhancement, etc. These maintenance tasks often result in change in road geometry features such as road grade~\cite{mahanpoor2019sustainable}. Remote sensing techniques, such as using satellite/aerial imagery data ~\cite{wang2017google}, can generate features of large road networks quickly. However, the high deployment cost makes these approaches impractical for coping with frequent data updates. What's more, they often produce low resolution/accuracy data~\cite{srtm} and are susceptible to occlusions by surrounding buildings, vegetation and shadows. Ground sensing techniques ~\cite{aran,jauch2018road}, which typically rely on specialized instrumented vehicles, can provide the desired accuracy. But, they also suffer from high deployment cost due to expensive equipment/sensors and dedicated labor, thus rendering them difficult to scale. An ideal sensing framework for the task of road grade estimation should be scalable, cost-efficient, and capable of providing frequent data updates. Also, the system should be accurate enough to enable the aforementioned applications such as ADAS and Autonomous driving.

Due to their ubiquity and rich array of onboard sensors (IMU, GPS and Magnetometer), smartphones provide a unique opportunity for developing a cost-efficient and scalable solution for road grade estimation. Assuming that a smartphone is stationary inside a moving vehicle, the 3D orientation estimates of pitch, yaw and roll of the smartphone can yield information about the grade of the road the vehicle is traveling on. Various techniques for real-time 3D orientation tracking of smartphones have been proposed \cite{A3, shen2018closing}, but are not directly suitable for our problem at hand, mainly due to incapability to cope with errors and biases of Smartphone's sensors in a dynamic driving environment.

In this paper, we propose a novel and easily deployable solution to estimate road grade using crowd-sourced data from smartphones. The proposed solution leverages acceleration and angular velocity data from accelerometer and gyroscope of the smartphone and vehicle speed data from OBD-II/GPS. We analyze performance of smartphone's gyroscope and accelerometer in a driving environment by characterizing the nature of error responses of both sensors. In particular, the study reveals that gyroscope is precise in capturing the shape of the road, but suffers from error accumulation/drift for estimation over long periods of time. Also, the drift is unpredictable in nature and can change over the course of a driving trip. On the other hand, accelerometer is not prone to drift, but is susceptible to large errors which are typically correlated to the dynamics of the vehicle. Based on observations derived from the study, we propose to use gyroscope as the primary sensor for road grade estimation. Accelerometer, on the other hand, is used to opportunistically provide chosen anchor snapshots, which are used to correct drift of gyroscope. Furthermore, we integrate Google elevation data~\cite{google} to compensate biases in grade estimations from accelerometer. Finally, we aggregate data from multiple sources to improve accuracy and robustness of the system by handling varied quality of sensor data from different sources.

The contribution of our work is three-fold. (1) We provide a comprehensive analysis on the performance of smartphone's gyroscope and accelerometer in a dynamic driving environment; (2) We develop a novel, practical, and easily deployable solution for road grade estimation using crowd-sourced data from smartphones; and (3) We extensively evaluate our proposed method on a $\approx$ 9km route, using natural driving traces from smartphones. The system achieves 90\% error less than 0.3$^\circ$ in estimating road grade and outperforms existing approaches by a considerable margin.

The rest of the paper is organized as follows. Sec.~\ref{background} presents the background and motivation of our problem. Sec.~\ref{acc_gyro_est_method} lays down the methodology of grade estimation using accelerometer and gyroscope and presents analysis on error characteristics of both sensors. Sec.~\ref{design} describes the design of the proposed road grade estimation technique. Sec.~\ref{eval} presents experimental evaluation results. Sec.~\ref{discuss} presents discussion and future work. Sec.~\ref{conclude} concludes the paper.

\section{Background and Motivation}\label{background}
A cost-efficient and scalable solution for estimation of road grade can provide a major upgrade to existing 2D digital maps. Current navigation service, that relies on 2D digital maps, can be made more robust using grade information, especially in complex environments such as multi-layered roads, by providing information on the elevation of the road the vehicle is on. Using road grade information, the current navigation service can be extended to suggest fuel-efficient routes in addition to the shortest~\cite{liu2010fuel}. Furthermore, gains in navigation accuracy can be achieved from improved localization techniques, especially in urban environments where accuracy of GPS is low~\cite{vemulapalli2011pitch}.

The benefits of information on road grade are not limited to improved navigation services. On the way towards autonomy, modern vehicles implement Advanced Driving Assistance Systems (ADAS)~\cite{adas} to help drivers with vehicle safety and efficiency. Having accurate information of road grade  enables a variety of ADAS applications. Examples include a) Predictive Powertrain Control System~\cite{powertrain}, which can achieve significant fuel savings by switching to optimal vehicle speed and transmission based on grade of the road ahead, and b) Curve Lightning and Beam Height Control System
, which can actuate motorized headlights to be turned up or down before the vehicle approaches an uphill or downhill.

In the future, fully autonomous operation of vehicles would require highly accurate information about the road network, which is provided by a High-Definition Map (or HD Map for short)~\cite{hdmaps}. A HD map is an accurate and highly attributed depiction of the road and can help an autonomous vehicle to precisely locate themselves and plan their path to the destination. Accurate information about road grade is an essential component of a HD map. Thus, an effective solution for estimation and collection of road grade will directly impact the autonomous industry by scaling creation of HD maps of the world-wide road networks. Furthermore, road grade information can be utilized to achieve improved accuracy in 3D-Object detection using Lidars for autonomous vehicles, especially over long range, where data from Lidars is sparse~\cite{yang2018hdnet}.

\begin{figure}
    \centering
    \begin{minipage} [t] {0.51\columnwidth}
        \centering
        \includegraphics[width=\textwidth]{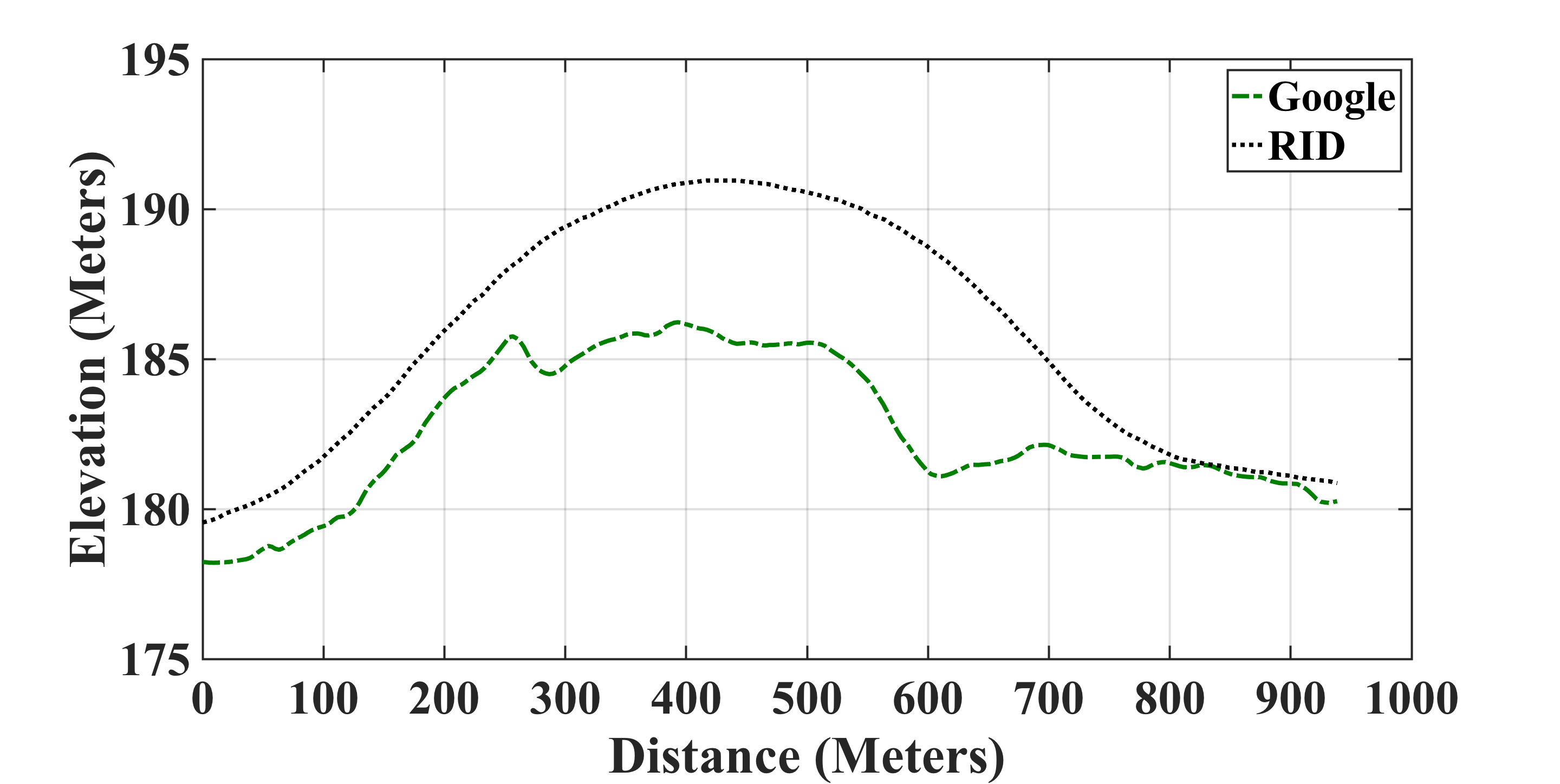} 
        \caption{Comparison of Google elevation data with RID benchmark.}
        \label{fig:elevation_comp}
    \end{minipage}\hfill
    \begin{minipage} [t] {0.45\columnwidth}
        \centering
        \includegraphics[width=\textwidth]{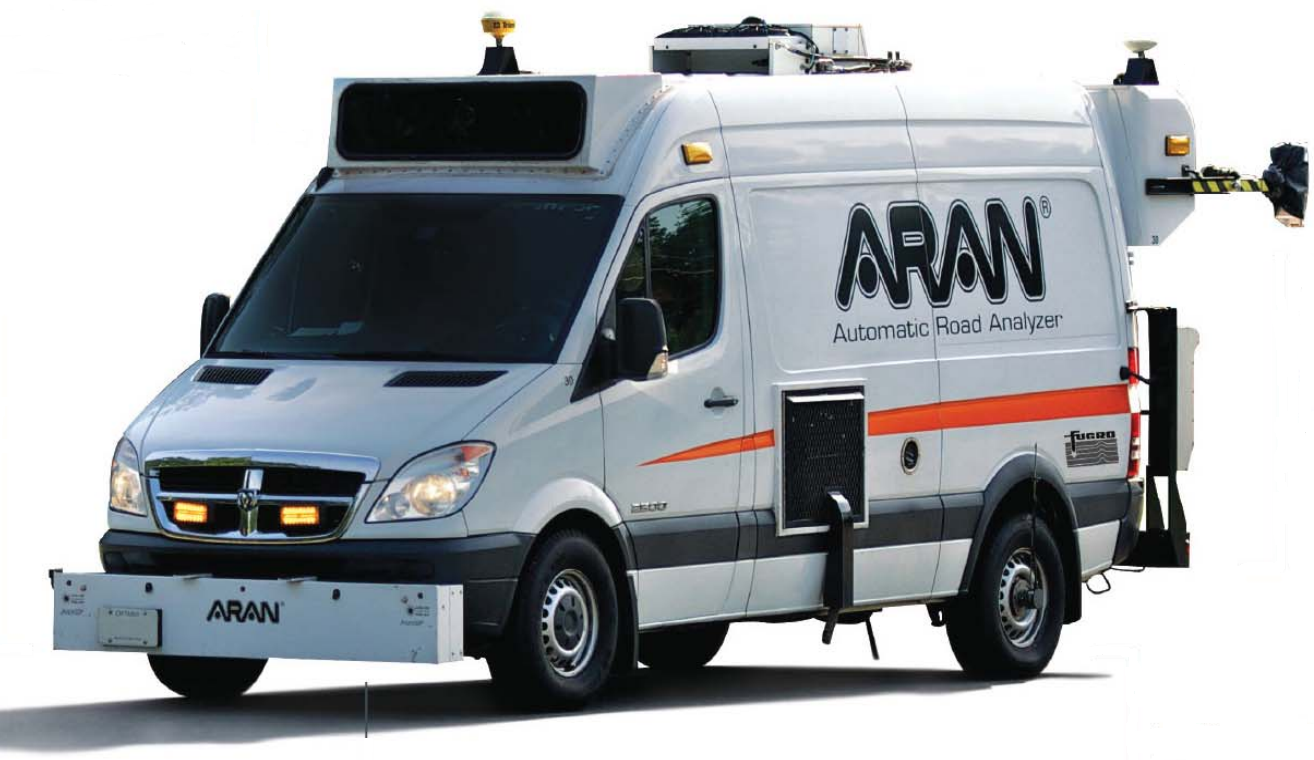} 
        \caption{Automatic Road Analyzer.}
        \label{fig:aran}
    \end{minipage}
\end{figure}

Traditionally, road grade is estimated using mainly remote sensing data (i.e., satellite or aerial imagery)~\cite{wang2017google}. Although these solutions are capable of providing large scale estimation of road geometry features, prohibitively high deployment and collection cost make these methods impractical for frequent data updates. Another major issue with these methods is their poor accuracy. To demonstrate this, we compare the samples of Google elevation data~\cite{google} (mainly derived from~\cite{srtm}) on a bridge, with elevation data from Road Inventory Database~\cite{rid}\footnote{Roadway Information Database includes information on road geometry features (Curvature, grade and superelevation) of $\sim$25,000 directional miles of roadway in six sites in USA. The data is collected using ARAN (Automatic Road Analyzer)~\cite{aran}, an instrumented vehicle with high-grade IMU's, laser scanners, high-precision GPS, and camera.}. As can be seen in Figure~\ref{fig:elevation_comp}, the elevation profile from Google is way off.



In addition to remote sensing based solutions, there are also ground sensing based solutions that employ specialized vehicles instrumented with high-quality sensors such as lidars, laser scanners, cameras and high-end IMU's for the task of road geometry estimation~\cite{aran,jauch2018road,zhang2013real}. For example, the aforementioned Road Inventory Database is collected using ARAN (Automatic Road Analyzer)~\cite{aran} as shown in Figure~\ref{fig:aran}. Though being able to achieve high accuracy, these approaches come with high deployment cost, and thus have limited scalability. To alleviate high cost,~\cite{kang2019road,sahlholm2011distributed} applies vehicle dynamics based model to estimate road grade leveraging data from vehicle's CAN bus and smartphone. However, due to the requirement of vehicle specific properties such as mass, frontal area, etc., the above methods lack generalizability and scalability.

Smartphones are an attractive sensing platform for estimation of road grade. It has an array of sensors, including IMU's, GPS, and magnetometer. Proliferation of smartphones due to cost reduction in recent years provides a unique opportunity to use it for large scale sensing tasks. For example,~\cite{mohan2008nericell,eriksson2008pothole} leverage smartphone's sensors for applications such as traffic and road surface condition monitoring. Our work also uses smartphone sensor data, but focuses on a different application, i.e. estimating accurate road grade profiles.

When fixed in vehicle, a smartphone can sense the orientation changes of the vehicle induced by road geometry. Compared to their professional counterparts, however, smartphone sensors are oftentimes of lower quality and accuracy, making it a challenging task to use them for road grade estimation as the acceptable margin of error is usually small. Commodity smartphones' MEMS 
sensors are prone to noise, biases, and drifts. For example, the bias error for the gyroscope in iPhone 4 is  $\pm1.15 \deg/\sec$~ \cite{kos2016suitability}. Furthermore, the biases can be dynamic in nature and thus change over time due to factors such as temperature and linear acceleration~\cite{A3}. Sensor fusion techniques such kalman and complimentary filters are typically applied to improve accuracy in a multi-sensor system~\cite{jauch2018road, zhang2013real}. However, it has been shown that directly adopting such techniques on smartphones produces sub-optimal results. Errors of 40$^\circ$ or higher have been reported in many cases using Android API, which directly adopts an implementation of Kalman filter~\cite{A3}. Furthermore, complex error characteristics of Smartphone's sensors in a dynamic driving environment make things even worse. As an example, Yang et al uses a complimentary filter to add/fuse estimate from accelerometer and magnetometer with estimate from gyroscope to calculate road grade~\cite{yang2016low}. However, as we explore in the evaluation, such fusion produces sub-optimal result predominately because of its inability to counter errors induced by unpredictable nature of gyroscope drift and the high accelerometer noise in dynamic driving conditions.

In theory, smartphone orientation estimation frameworks, which intelligently incorporate readings from gyroscope and other IMU's can be used for road grade estimation~\cite{A3, shen2018closing}. However, direct adoption of these techniques is not suitable for our application due to the following reasons. a) These solutions rely on static or pure rotational moments (estimate accurate gravity) to opportunistically calibrate the orientation estimates. Such opportunities are plenty in daily activities such as walking, eating, etc. However, a smartphone kept inside the car is constantly acted upon by forces due to dynamics of the vehicle and thus such opportunities will be rare when the vehicle is in motion. b) These solutions rely on unpolluted magnetometer readings for desired accuracy. Inside a vehicle, the magnetometer is susceptible to constant biases and dynamic noise due to presence of strong ferromagnetic materials and electronic devices in vicinity
. It has been reported that the accuracy of magnetometer (and thus the corresponding solution systems) degrades drastically, when operating in environments where magnetic interference is strong~\cite{A3, shen2018closing}. Typically, these biases can be dealt with by magnetic field profiling~\cite{yang2016low}. This additional step, however, is undesirable for a smartphone-based crowd-sourcing system, as it might discourage the participants to perform the actual tasks. Therefore to make our system transparent to end users, we do not leverage magnetometer in our proposed framework, and instead rely on gyroscope and accelerometer as the primary sensors.

\section{Road grade estimation preliminaries}\label{acc_gyro_est_method}

The IMU sensors on a smartphone fixed inside a moving vehicle will be able to capture the dynamic of the vehicle, which is typically a result of the forces induced by driver control (acceleration/braking done to achieve desired speed and steering control for lane changes on a straight road), as well as the road's horizontal (steering control while negotiating a turn) and vertical (while going up-hill or down-hill a road/while travelling on a road with super-elevation) geometry. Therefore, if the contribution of driver control is removed from the smartphone signal, we will be left with a signal encapsulating road geometry information. The proposed approach for road grade estimation from a smartphone is based on the above intuition. In this section, we will present the integrated components of our design.

\subsection{Data Processing}\label{data_process}
\subsubsection{Preprocessing}\label{data_alig}
Due to their relatively low quality, smartphone sensors tend to output data more prone to noise, which is further amplified by vibrations of the vehicle. Therefore, we smooth the signal from accelerometer and gyroscope by passing it through a second-order Butterworth low-pass filter. The accelerometer and gyroscope are sampled at 200Hz, whereas the velocity data from OBD-II and GPS arrives at a much lower rate of 10Hz and 1Hz, respectively. To perform trace alignment, we thus interpolate velocity data to get 200 samples/sec. We also employ trace synchronization~\cite{wang2013sensing} to prevent data from the Smartphone and the OBD-II scanner to go out of sync. Finally, we segregate data for different road segments on the test route (Fig.~\ref{fig:test_route}). We divide the test route into road segments based on presence of an intersection or a stop sign.

\subsubsection{Coordinate Alignment}\label{coor_align}
\begin{figure}[htbp]
	\centering
	\includegraphics[width=0.7\linewidth]{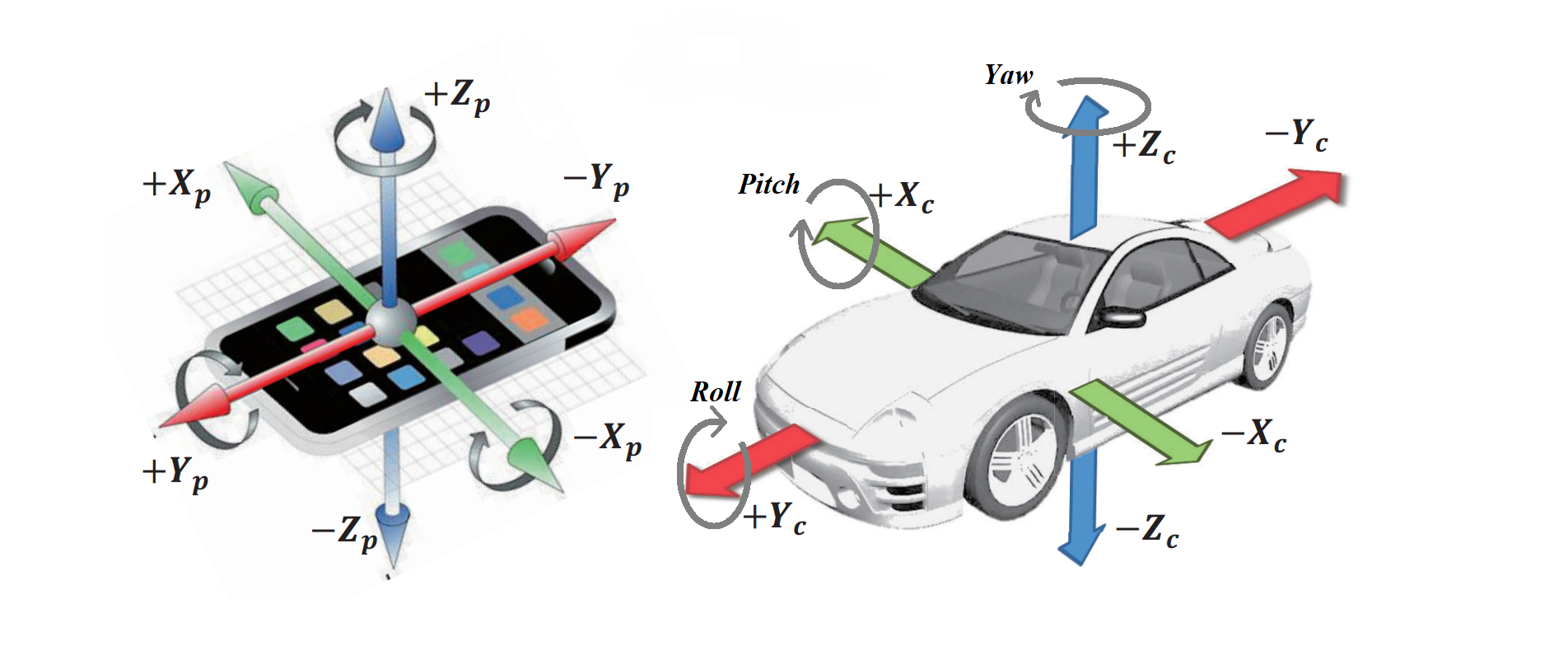}
	\caption{Smartphone and Vehicle Coordinate System.}
	\label{fig:coord}
\end{figure}

To sense meaningful dynamics of the vehicle using a smartphone, it is necessary to align the phone's coordinate system with the vehicle's. We will work with the coordinate system shown in Fig.~\ref{fig:coord} in this paper. We leverage an existing technique to perform the alignment~\cite{wang2013sensing}, which results in a $3\times 3$ rotation matrix $R_{PC}$ for transforming the smartphone's data to the vehicle's frame of reference. The first, second and third columns of  $R_{PC}$ are unit vectors $\vec{x_u}$, $\vec{y_u}$ and $\vec{z_u}$ pointing in the direction $+X_C$, $+Y_C$ and $+Z_C$ respectively as also illustrated in Fig.~\ref{fig:coord}. $\vec{z_u}$ is estimated when the vehicle is stationary and thus, the acceleration reported by the accelerometer includes gravity component. We estimate $\vec{z_u}$ at the beginning of a trip, when the vehicle is on a level surface (e.g., in a parking lot).  $\vec{y_u}$, which points in the moving direction of the vehicle, is estimated when the vehicle undergoes acceleration on a straight road segment (e.g., when the vehicle starts from zero velocity after stopping at an intersection). Finally, $\vec{x_u}$ is the cross product of $\vec{y_u}$ and $\vec{z_u}$. Coordinate alignment is done once, during a trip, as soon the system captures a valid acceleration profile for the estimation of $\vec{y_u}$. We also use {\em pitch}, {\em yaw}, and {\em roll}, to denote the vehicle's rotations about the lateral axis $X_C$, perpendicular axis $Y_C$, and longitudinal axis $Z_C$, respectively.


\subsection{Road Grade Estimation Framework }\label{grade_est}

Varying 3D road geometry results in rotation of vehicle in the vertical $Y_C$-$Z_C$ and $X_C$-$Z_C$ planes (Fig.~\ref{fig:coord}). The smartphone fixed in the vehicle also undergoes changes in orientation accordingly. Specifically, varying road grade will result in rotation of the vehicle about the $X_C$ axis, thus contributing to the change in the vehicle's pitch. We discuss our multimodal estimation methodologies as follows.\\

\subsubsection{Pitch Estimation Using Gyroscope}\label{pitch_gyro}

A smartphone's gyroscope reports the real-time angular velocities $\omega_{X_p,t}$, $\omega_{Y_p,t}$, and $\omega_{Z_p,t}$ around the $X_p$, $Y_p$ and $Z_p$ axis of the Smartphone, respectively. Assuming that gyroscope data has been aligned with vehicles coordinate frame, the pitch at time instant t ($\theta_{gyro,t}$) can be estimated by integrating the angular velocity of the vehicle about $X_C$ axis ($\omega_{X_c,t}$) using Eq.~\ref{eq:pitch_gyro}.
	
	\begin{equation} \label{eq:pitch_gyro}
	\theta_{gyro,t} = \theta_{gyro,t-1} + \omega_{X_c,t}.\varDelta{t}
	\end{equation}
	
	\begin{figure}[htbp]
		\centering
		\includegraphics[width=0.8\linewidth]{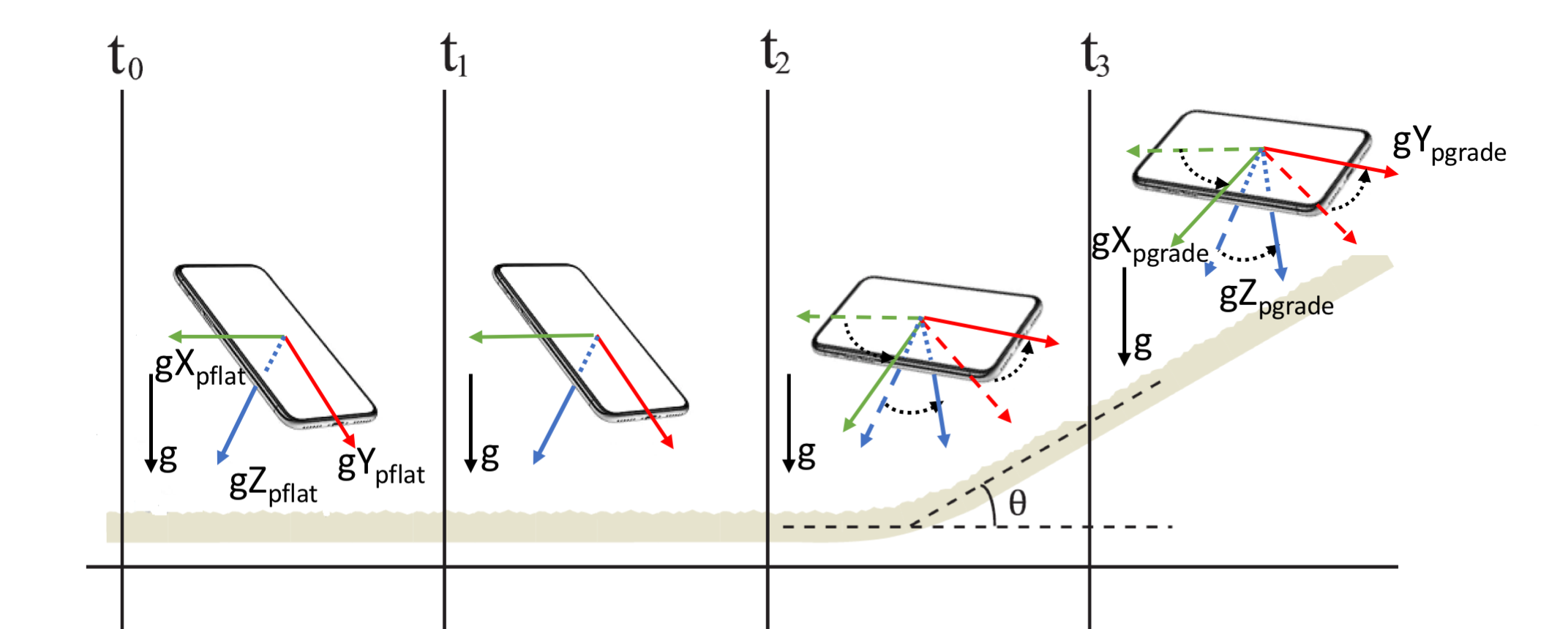}
		\caption{Redistribution of gravity components along smartphone's axes ($X_p$, $Y_p$ and $Z_p$), when vehicle moves from a level road segment to a segment with grade.}
		\label{fig:grav_rot}
	\end{figure}
	
\subsubsection{Pitch Estimation Using Accelerometer}\label{pitch_acc}
The acceleration captured by a smartphone's accelerometer contains both the vehicle's longitudinal/lateral accelerations and the gravity. When a vehicle is stationary, the measured acceleration of the smartphone is only due to gravity. When the vehicle is moving, road grade will result in rotation of vehicle around $X_c$, contributing to pitch of the vehicle. This will result in rotation of the gravity vector. As illustrated in Fig.~\ref{fig:grav_rot}, traveling on a road segment with grade will result in rotation of the smartphone, along with the vehicle. This will lead to redistribution of gravity vector components along smartphone's axes. While traveling on a level road, gravity experienced by smartphone is decomposed into $gX_{pflat}$, $gY_{pflat}$ and $gZ_{pflat}$, along $X_{p}$, $Y_{p}$ and $Z_{p}$ respectively. While traveling on a road segment with grade, redistribution of gravity results into components $gX_{pgrade}$, $gY_{pgrade}$ and $gZ_{pgrade}$, along $X_{p}$, $Y_{p}$ and $Z_{p}$ respectively. Furthermore, moving direction $\vec{y_u}$ remains constant during trip. Therefore, pitch of the vehicle can be estimated by tracking gravity vector's rotation in the plane determined by $\vec{y_u}$ and $\vec{z_u}$, where $\vec{z_u}$ is estimated when the vehicle is stationary on a level surface during coordinate alignment process.\\
	
However, when the vehicle is moving, accelerometer captures longitudinal (braking and accelerating of the vehicle) and lateral acceleration (when the vehicle is negotiating a turn/lane change), along with gravity. Therefore, to estimate gravity, contribution of longitudinal and lateral acceleration has to be removed from accelerometer readings. We leverage speed information of a vehicle from OBD-II port of the vehicle or the GPS device on the smartphone to estimate longitudinal and lateral acceleration of the vehicle. Longitudinal acceleration magnitude ($A_t$) of the vehicle is estimated by differentiating the speed and projecting it on $Y_C$ using Eq.~\ref{eq:lon_acc}. Lateral acceleration of the vehicle is estimated using velocity $V_t$ and angular velocity about $Z_c$ ($\omega_{Z_c,t}$) using Eq.~\ref{eq:lat_acc}.
	
	\begin{equation} \label{eq:lon_acc}
	\vec{A_{Y_c,t}} =A_t.\vec{y_u}
	\end{equation}
	
	\begin{equation} \label{eq:lat_acc}
	\vec{A_{X_c,t}} =( \omega_{Z_c,t}.V_t).\vec{x_u}
	\end{equation}
	
When the vehicle is moving, the gravity vector $\vec{G_{t}}$ can be estimated using Eq. \ref{eq:grav_t} where, $\vec{A_t}$ is the acceleration vector reported by the smartphone, $\vec{A_{Y_c,t}}$ and $\vec{A_{X_c,t}}$ are longitudinal and lateral acceleration of the vehicle.
	
	\begin{equation} \label{eq:grav_t}
	\vec{G_{t}} = \vec{A_t} - \vec{A_{Y_c,t}} - \vec{A_{X_c,t}} 
	\end{equation}
	
Finally, pitch estimation using accelerometer ($  \theta_{acc,t}$) is done by tracking rotation of gravity vector in the plane determined by $\vec{y_u}$ and $\vec{z_u}$ (Eq.\ref{eq:pitch_acc}). Fig. \ref{fig:acc-gyro-est-grade_a} illustrates the estimated road grade (labeled '' Grade-Acc '') using the approach described above for an example road segment.

	\begin{equation} \label{eq:pitch_acc}
	\theta_{acc,t} = \angle (\vec{G_{t}}, \vec{y_{u}})
	\end{equation}


\subsection{Understanding the nature of Gyroscope and Accelerometer} \label{nature_gyro_acc}

To analyze road grade estimation using the approaches described above, we performed an analysis on responses of gyroscope and accelerometer. Data used for the analysis is comprised of 3 trips/loops on our test route depicted in Fig.~\ref{fig:test_route}). Also, data collection was done in a single continuous session without stoppages in between trips. The smartphone was secured on the windshield of the vehicle using a phone holder for the entire data collection period (about 50 mins of driving). The vehicle was kept stationary for 30 seconds at the beginning  of the trip for coordinate alignment. For gyroscope-based estimation, we initialize the road grade with the ground-truth value at the start of each road segment and integrate the successive samples.


\begin{figure*}
	\centering
	\begin{subfigure} [t] {0.33\textwidth}
		\includegraphics[width=0.97\textwidth]{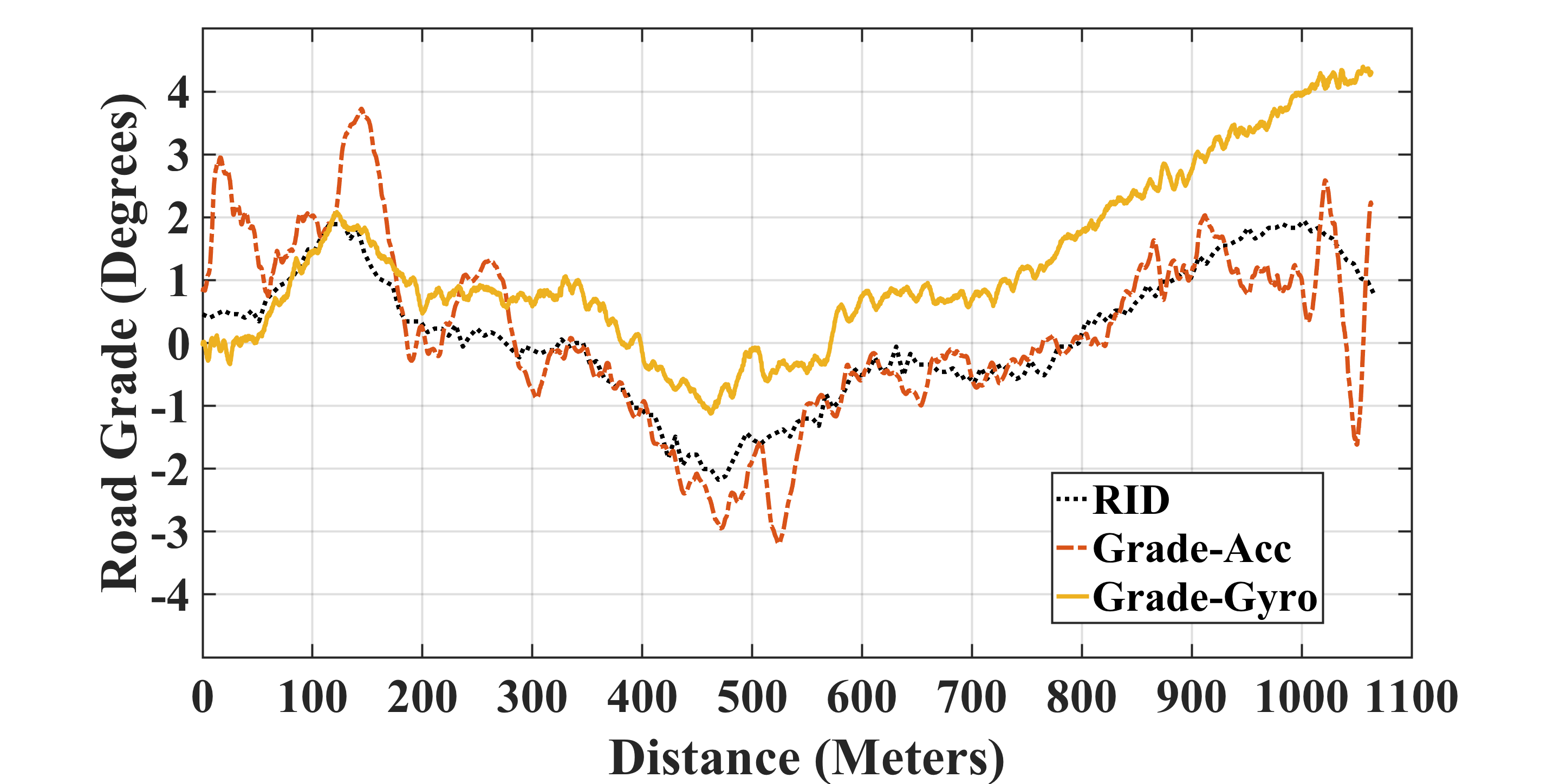}
		\caption{Estimate of road grade using gyroscope and accelerometer.}
		\label{fig:acc-gyro-est-grade_a}
	\end{subfigure}
	\begin{subfigure}[t] {0.33\textwidth}
		\includegraphics[width=0.99\textwidth]{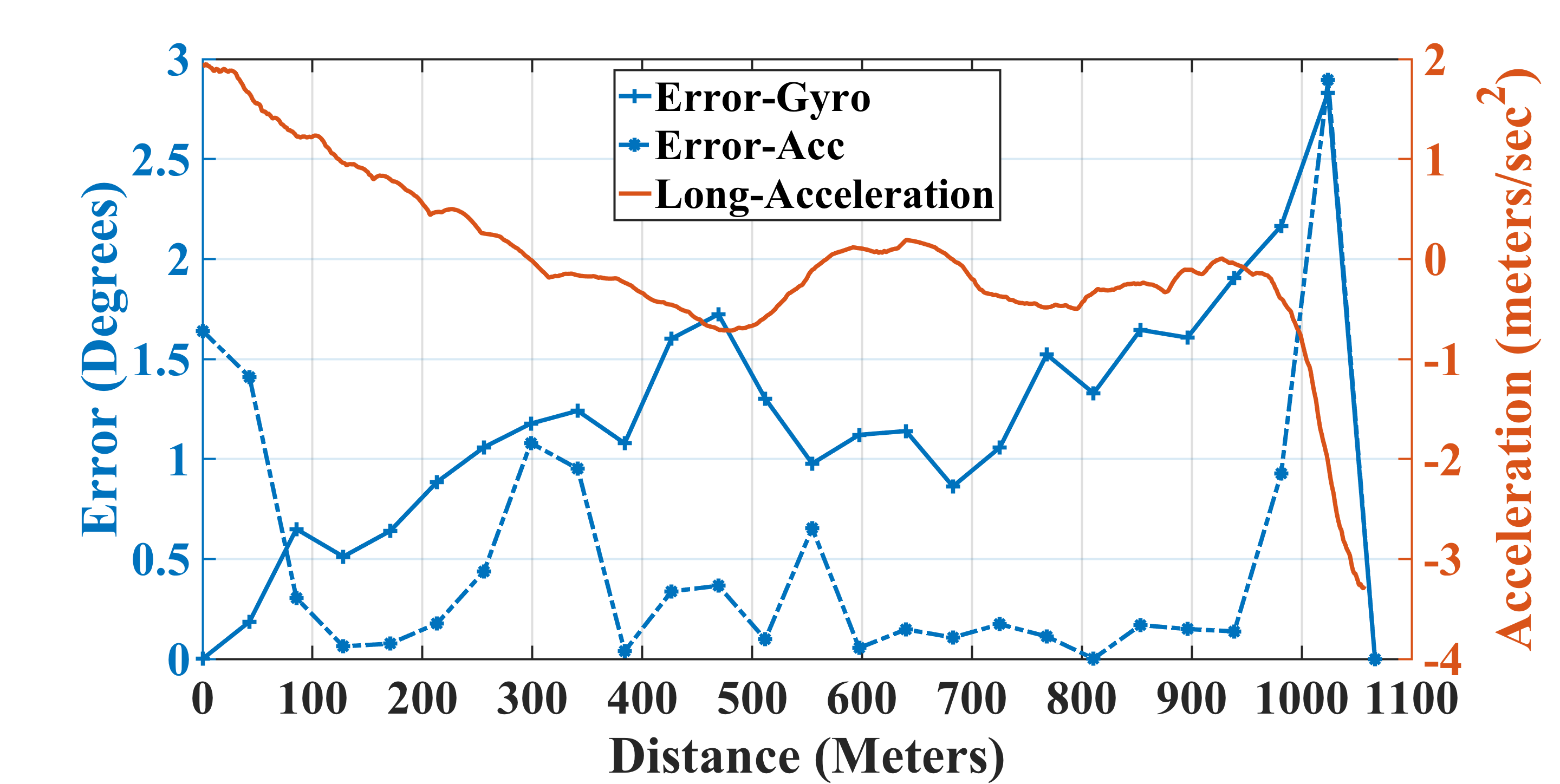}
		\caption{Error characteristics of accelerometer and gyroscope in estimation of road grade.}
		\label{fig:error_anal}
	\end{subfigure}
	\begin{subfigure}[t] {0.33\textwidth}
		\includegraphics[width=\textwidth]{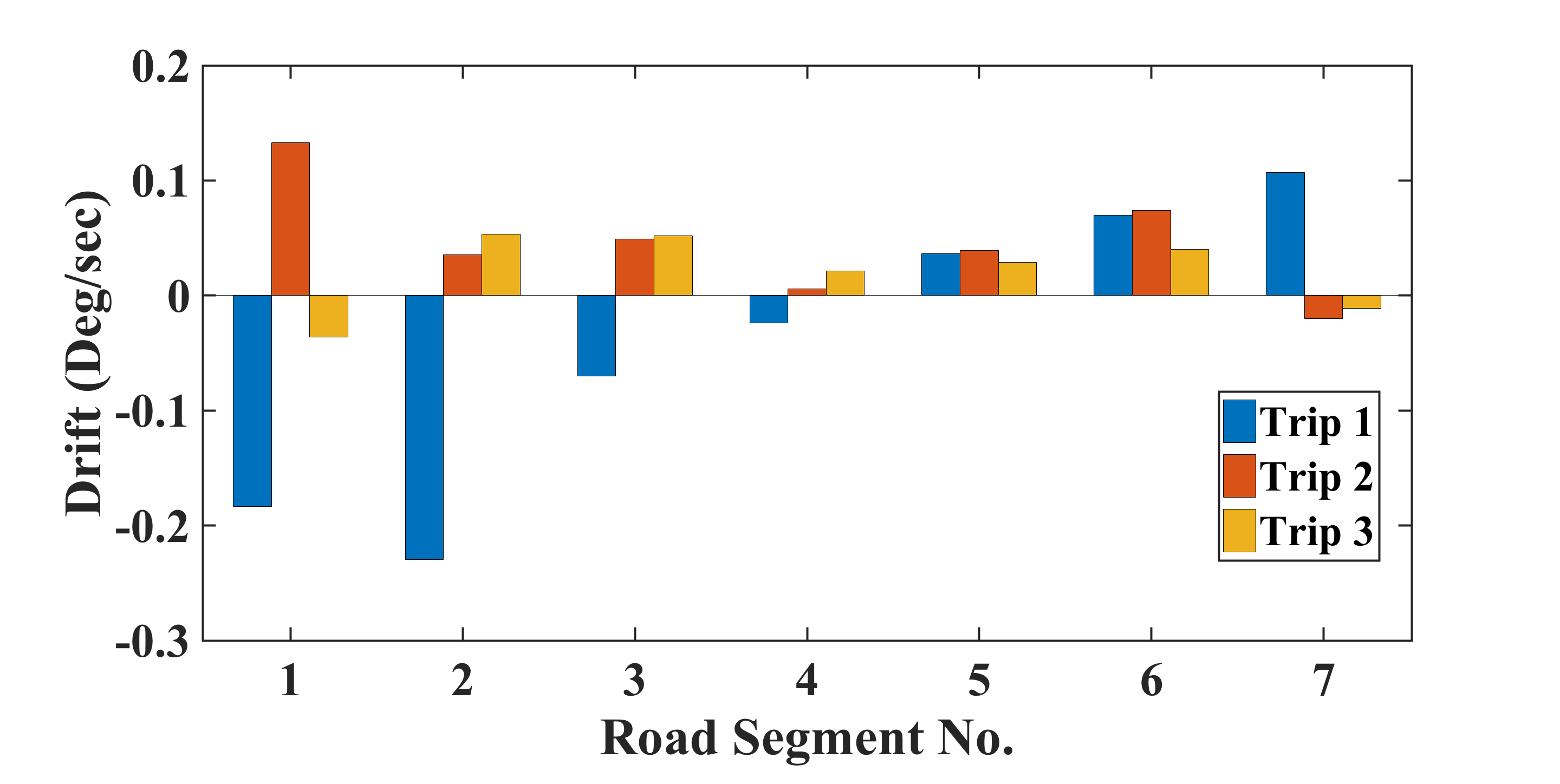}
		\caption{Analysis of gyroscope drift.}
		\label{fig:drift_anal}
	\end{subfigure}
	\caption{Analysis of road grade estimation using gyroscope and accelerometer.}
	\label{fig:analysis_acc_gyro}
\end{figure*}

Fig.\ref{fig:acc-gyro-est-grade_a} illustrates the response of gyroscope and accelerometer for road grade estimation for a road segment on the route. The error accumulation in gyroscope's response is evident from the result, as estimation drifts from the ground-truth with time. However, grade estimated using gyroscope is precise in capturing the shape of the profiled road segments. On the other hand, the response of accelerometer is not prone to drift. But, the estimation is susceptible to large errors with high variance. Fig.~\ref{fig:error_anal} indicates that estimation using accelerometer is prone to large errors when the vehicle dynamics are not stable. Specifically, we observe peaks in error when a) vehicle undergoes strong acceleration/deceleration, b) the rate of change of acceleration/deceleration, defined as ``jerk'', is rapid. In comparison, estimation using gyroscope is not susceptible to varying vehicle dynamics and drift is its major source of error.


Fig. \ref{fig:drift_anal} plots the drift of gyroscope while estimating road grade on different road segments across 3 trips. The unpredictable nature of drift is evident from the figure. Firstly, the drifts can be as large as $\approx$-0.2$^\circ$/sec (Trip 1 and segment 2). Secondly, the drifts not only change with time over the course of a trip, but also show significant variation, both in magnitude and direction, on the same road segment across different trips. For example, the drift of gyroscope for trip 1 is $\approx$-0.18$^\circ$/sec for road segment 1 and is $\approx$0.1$^\circ$/sec for road segment 7. For road segment 1, the drift varies from  $\approx$0.18$^\circ$/sec to $\approx$0.13$^\circ$/sec. The unpredictability of gyroscope drifts is partly attributed to changing road surface conditions. Varying vibration profiles of the vehicle on roads with different surface conditions lead to different noise levels, thus causing dynamic drifts. Furthermore, imperfect calibration/coordinate alignment can result in errors due to mixing of gyroscope signals across axes. For example, pitch estimation might get distorted by yaw values, when the vehicle encounters a turn. Such errors are significant in the context of road grade estimation, where high precision is needed for applications such as ADAS.



The preliminary analysis indicates the ``complimentary'' nature of gyroscope and accelerometer w.r.t. to the task of road grade estimation, making a strong case for fusing both sensors for improving the estimation accuracy.

\begin{figure}[htbp]
	\centering
	\includegraphics[width=0.8\linewidth]{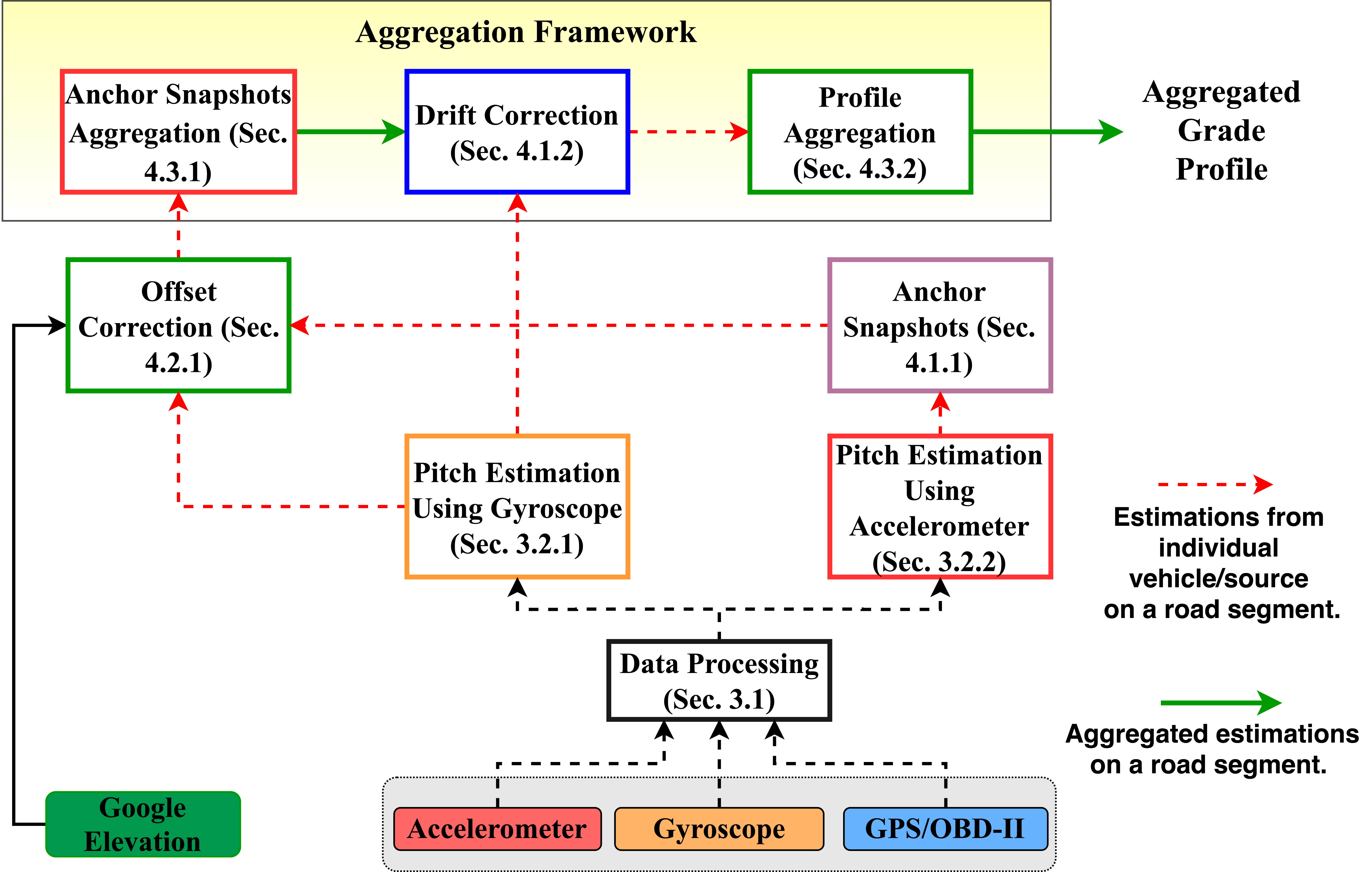}
	\caption{System Architecture. }
	\label{fig:sys_arch}
\end{figure}
\section{Design} \label{design}

Deriving insights from our understanding of error characteristics of smartphone's sensors in a dynamic driving environment, we propose the design of our road grade estimation framework. Fig. \ref{fig:sys_arch} illustrates the architecture of our proposed system for road grade estimation. 
The raw data from the smartphone is pre-processed in the ``Data Processing'' module, as described in Sec.~\ref{data_process}. Estimates of vehicle's pitch are computed using data from gyroscope and accelerometer in ``Pitch Estimation Using Gyroscope'' and ``Pitch Estimation Using Accelerometer'' modules, respectively, as described in Sec. \ref{grade_est}. The ``Anchor Snapshots'' module opportunistically selects grade estimation during stable driving phases, to handle error prone estimations from accelerometer due to unstable vehicle dynamics, as described in Sec.~\ref{snaps}. The bias associated with the computed anchor snapshots is handled in the ``Offset Correction'' module, by leveraging information from Google elevation data and estimations from gyroscope, as described in Sec. \ref{goog_elev}. The ``Anchor Snapshots Aggregation'' module aggregates anchor snapshots from different vehicles/sources on a road segment, as described in Sec.~\ref{agg_snaps}. The aggregated snapshots are then fused with the pitch estimations from gyroscope in the ``Drift Correction'' module, to generate drift compensated grade profiles of a road segment, as described in Sec.~\ref{drift_corr}. These drift compensated grade profiles are then aggregated in the ``Profile Aggregation'' module to estimate a single road grade profile of a road segment, as described in Sec.~\ref{agg_prof}.


\subsection{Opportunistic Road Grade estimation} \label{oppor_est}
As explored in the previous section, gyroscope and accelerometer have complimentary properties w.r.t the task of grade road  estimation. Traditional sensor fusion techniques such as kalman and complimentary filters can exploit sensor redundancy to improve accuracy. For optimal performance, these fusion techniques require estimates of sensor's noise/error characteristics, which are fed as parameters to the fusion framework. However, as shown in the previous section, the quality of smartphone sensor's response can be unpredictable in nature. Thus, in dynamic environments, it is difficult to optimize parameters for such fusion techniques to perform optimal error control. Therefore, directly adopting such sensor fusion approaches for the task of road grade estimation produces sub-optimal results, as demonstrated in the Sec. \ref{eval}.

Based on our understanding of smartphone sensor's quality in different conditions, we propose an opportunistic framework for road grade estimation, with the underlying intuition as follows.

\begin{itemize}
	\item We treat gyroscope as the ``primary'' sensor for the task of road grade estimation. This is based on the observation that estimation using gyroscope is precise in capturing the 3D shape of the road segment accurately. Also, the estimation is not prone to errors induced by varying vehicle dynamics.
	
	\item To compensate for the drift-induced errors in estimation from gyroscope, we leverage estimation from accelerometer as an anchor for correcting the drift. Specifically, we opportunistically select ``anchor snapshots'' during ``stable'' driving phases to correct drift of the signal from gyroscope.
	
	\item To counter unpredictable response of smartphone's gyroscope, drift correction is done ``on the go''. Specifically, we perform drift correction in windows of given distances.
\end{itemize}

Next we present our proposed framework in more detail.

\subsubsection{Capturing Anchor Snapshots} \label{snaps}

As previously discussed, grade estimation using accelerometer is error prone due to forces induced by vehicle dynamics. Constant or frequent acceleration/braking and turning/lane changes are the cause of these forces. Even if we remove the contribution of vehicle dynamics, as described in Sec.~\ref{pitch_acc}, the method is still not capable of fully compensating for the contribution of vehicle dynamics because of the noise in OBD-II/GPS data, thus resulting in inaccurate estimation of longitudinal and lateral acceleration of the vehicle. Furthermore, ideally OBD-II/GPS trace should be ``perfectly'' synchronised with the smartphone data trace, which we observed was difficult to obtain. Also, during strong acceleration/deceleration, the vehicle rotates around its lateral axis ($X_c$), the extent of which is dependent on the magnitude of acceleration and suspension properties of the vehicle~\cite{sahlholm2011distributed}. This phenomenon can be commonly observed as ``lifting up'' of the vehicle's front when it is accelerated from zero velocity (e.g., when starting off from a stand-still at an intersection) and could thus contribute to the pitch of vehicle. Because of the above, the error of grade estimation using accelerometer is magnified when the dynamics of vehicle are not stable, especially during rapid acceleration/deceleration phases, as can be observed in Fig.~\ref{fig:error_anal}.

Keeping the above in mind, we opportunistically filter out the grade estimations during stable driving phases based on the following two metrics~(Fig.~\ref{fig:sys_arch}, middle):

\begin{itemize}
	\item {\bf Acceleration}: We reject the estimations of road grade during ``high'' acceleration/deceleration phases. These phases are filtered out based on ${acc_{thresh}}$. 
	
	\item {\bf Jerk}: We also reject the estimations when the dynamics of the vehicle are unstable (e.g., rapid acceleration/deceleration phases). This is achieved by setting a ${jerk_{thresh}}$, where jerk is the rate of change of acceleration. 
\end{itemize}

The grade estimations during stable driving phases are segregated into bins of length 2m. Finally, observations in each bin are averaged to estimate the anchor snapshots. Fig.~\ref{fig:oppor_correction_grade_a} illustrates the estimated anchor snapshots using the approach described above. We perform an analysis of impact of ${acc_{thresh}}$ and ${jerk_{thresh}}$ on performance of our system in Sec.~\ref{eval}.

\subsubsection{On the Go Drift Correction} \label{drift_corr}

The anchor snapshots estimated in the previous section are used to estimate the gyroscope drift~(Fig.~\ref{fig:sys_arch}, top). We use the least-squares~\cite{york1966least} method to fit a line to the difference between the grade estimation from gyroscope and the anchor snapshots. To account for the dynamic nature of gyroscope drift, the correction is done in windows of distances of size ${wind_{drift}}$, as shown in Eq.~\ref{eq:wind_size}, where $R_{L}$ is the length of the profiled road segment and $D_{anch}$ is the list of distances between consecutive anchor points on the road segment. 

\begin{equation} \label{eq:wind_size}
wind_{drift} = \max((R_{L}/3),\max(D_{anch}))
\end{equation}

\begin{figure}[htbp]
	\centering
	\includegraphics[width=0.7\linewidth]{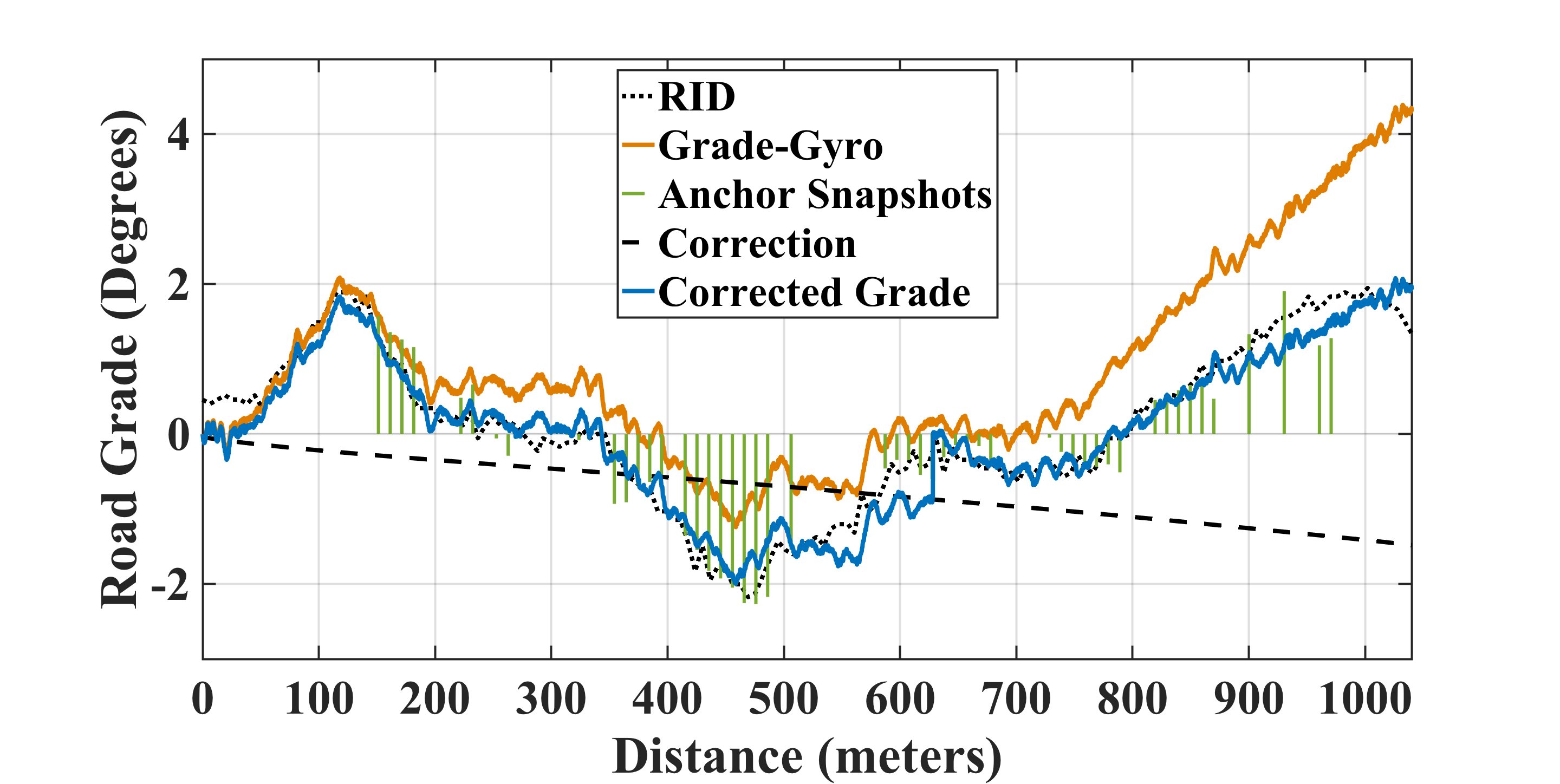}
	\caption{On the Go drift correction using anchor snapshots. }
	\label{fig:oppor_correction_grade_a}
\end{figure}

Fig.~\ref{fig:oppor_correction_grade_a} illustrates the result of on the go drift correction mechanism. The effectiveness of drift correction depends on a) Accuracy of anchor snapshots: the more accurate the anchor snapshots, the better the estimations of the drift; b) The density of anchor snapshots (the number of anchor snapshots per unit distance): Higher density anchor snapshots will provide more information for the drift correction mechanism to work with.

\subsection{The Offset Problem}\label{off_prob}
\begin{figure*}
	\centering
	\begin{subfigure}[t]  {0.33\textwidth}
		\includegraphics[width=\textwidth]{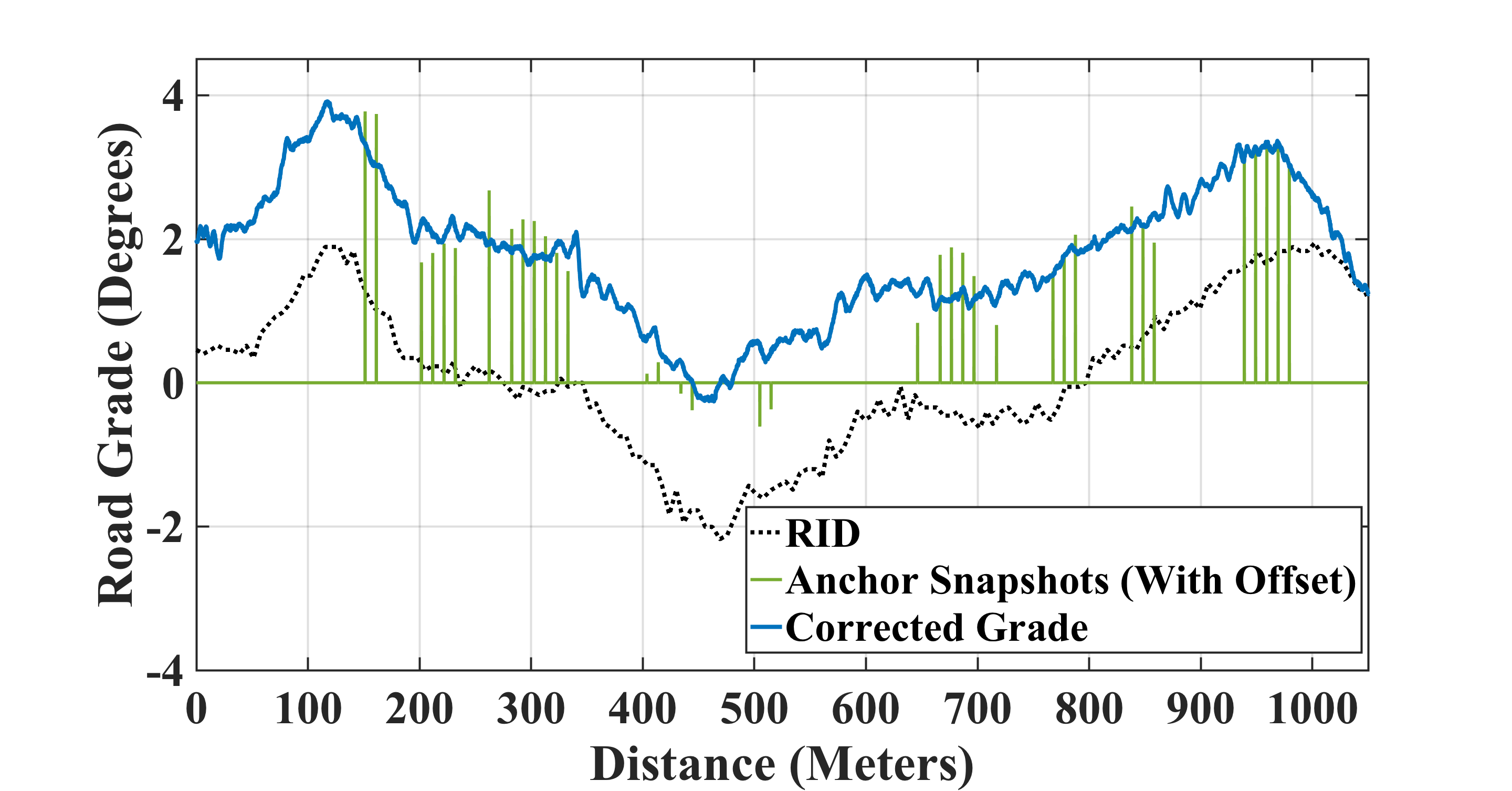}
		\caption{Offset in estimation of road grade.}
		\label{fig:offset_prob}
	\end{subfigure}
	\begin{subfigure}[t] {0.325\textwidth}
		\includegraphics[width=\textwidth]{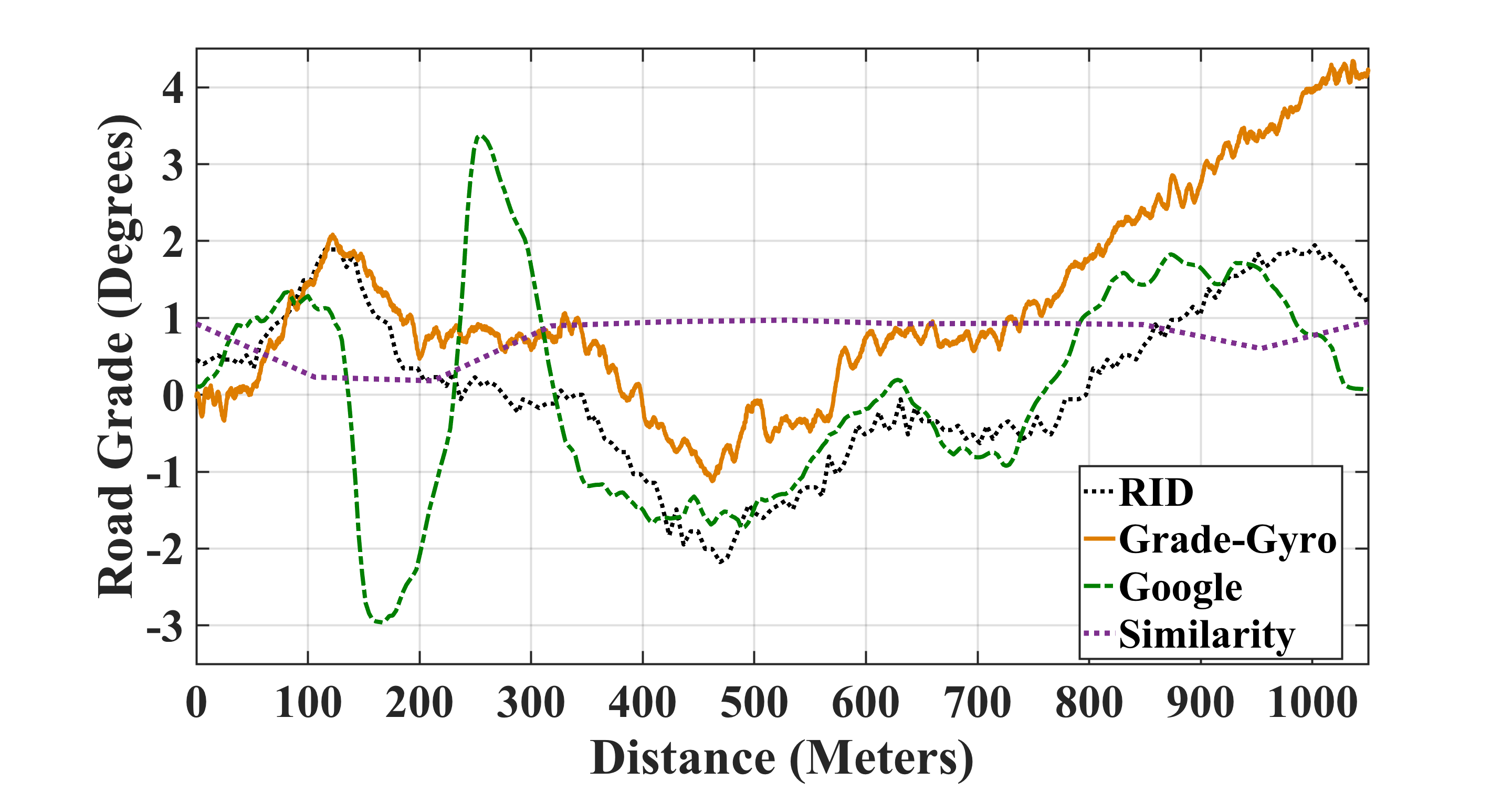}
		\caption{Shape similarity between grade estimation from gyroscope and Google elevation data.}
		\label{fig:similar}
	\end{subfigure}
	\begin{subfigure}[t] {0.33\textwidth}
		\includegraphics[width=\textwidth]{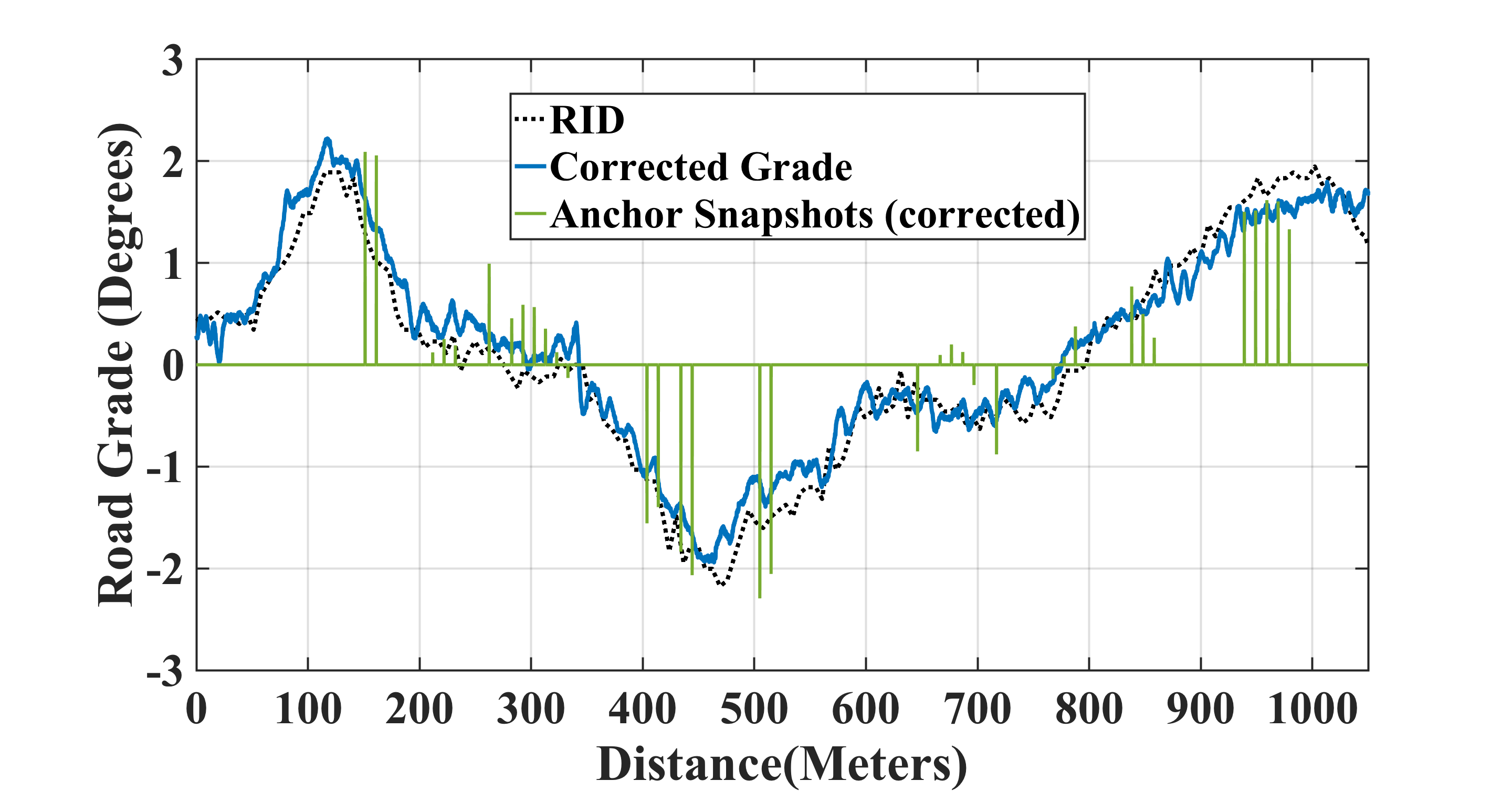}
		\caption{Road grade estimate after offset correction.}
		\label{fig:offsetted_out}
	\end{subfigure}
	\caption{Offset correction methodology using Google elevation data.}
	\label{fig:offset_correction}
\end{figure*}

The anchor snapshots as computed in Sec.~\ref{snaps} are characterized by an offset. As illustrated in Fig.~\ref{fig:offset_prob}, estimations of anchor snapshots exhibit a ''shift'' from the actual grade. The problem of offset is primarily attributed to :
{\bf a) Inaccurate estimation of vehicle's forward moving direction ($\vec{y_u}$) during calibration}: As mentioned in Sec.~\ref{snaps}, while undergoing strong acceleration/deceleration, a vehicle's body rotates about its lateral axis ($X_c$), which could result in inaccurate estimation of $\vec{y_u}$. {\bf b) Inaccurate estimation of the reference gravity vector ($\vec{z_u}$) during calibration}: Estimation of $\vec{z_u}$ on level road surface might not always be possible. While calculating $\vec{z_u}$ when the vehicle is stationary the road surface more often than not will have some inclination due to grade or cross-slope or both. 

The above result in errors that propagate through our accelerometer-based pitch estimation framework and manifest as an offset in the final road grade estimation. Fig. \ref{fig:offset_prob} demonstrates the effect of offset in estimation of anchor snapshots. The figure illustrates that the methodology described in Sec. \ref{snaps} and \ref{drift_corr} is able to compensate for the drift of the gyroscope. However, the estimation suffers from an offset of $\approx$ 2$^\circ$, which results in inaccurate estimation.


\subsubsection{Making use of Google Elevation Data}\label{goog_elev}

Remotely sensed data from the SRTM (Shuttle Radar Topography Mission)~\cite{srtm} is the major source of Google elevation data. We use Eq. \ref{eq:grade_google} to estimate grade from the elevation samples, where $E_{D}$ is the elevation gain over distance D.

\begin{equation} \label{eq:grade_google}
\theta_{google,d} = \arcsin\frac{E_{D}}{D}
\end{equation}

Google elevation data has the following general characteristics: 

\begin{itemize}
\item {\bf High Coverage: } SRTM data was collected using a specially modified radar system mounted on a space shuttle and covers 80\% of earth's landmass.

\item {\bf Low Overall Accuracy: } The Data has an absolute error up to 16m in elevation. Also, proneness to occlusions due to surrounding infrastructure and foliage affects the accuracy.

\item {\bf Low Resolution: } The data has spatial sampling of 30m $\times$ 30m and thus has in-sufficient resolution to capture an accurate shape of the road elevation profile to enable applications such as ADAS~\cite{adas}.

\end{itemize}


Though not good enough to provide the desired accuracy for road grade estimation in itself, Google elevation data can act as a good source of information to ``fix'' the offset described in Sec.~\ref{off_prob}. We observe that point estimates of road grade using Google elevation data is reasonably accurate when: {\bf a) Road segments follow elevation profile of the surrounding terrain}. In these cases, the impact of ``low resolution'' data from Google elevation in estimating road grade is low, as successive elevation samples capture the grade profile of the road even if they do not fall on the road segment. For road segments that do not follow the terrain (e.g., elevated structures such as bridges, ramps etc), the accuracy of grade estimates using Google elevation data degrades. {\bf b) Road segments are not interfered by occlusions.} 

In Fig. \ref{fig:similar}, since the road follows its surrounding terrain, estimates from Google elevation data is accurate from $\approx$ 400m to $\approx$ 800m. In contrast, the accuracy suffers from $\approx$ 50m to $\approx$ 300m because to occlusion due to an over-pass that goes over the road segment. Therefore, estimations from Google elevation data in selected regions of road can be utilized to estimate the offset of anchor snapshots. 

To extract regions of high accuracy of grade estimation from Google elevation data, we work with the following observations:  a) A region with grade profile estimated using google elevation data that captures the shape of the road accurately will indicate high accuracy of point estimates  in this region; and b) As observed in Sec. \ref{nature_gyro_acc}, grade estimations from gyroscope capture the shape of road accurately. Therefore, we compute the ``shape similarity'' between grade estimations from Google elevation data and the gyroscope to extract regions of high accuracy~(Fig. \ref{fig:sys_arch}, left). Suppose, $P = \{p_{x1}, p_{x2}, p_{x3}, \ldots, p_{xn}\}$ and $Q = \{q_{x1}, q_{x2}, q_{x3}, \ldots, q_{xn}\}$ are samples of grade estimated using Google elevation data and gyroscope, respectively, in detection window of length $d_{sim}$. Similarity between $P$ and $Q$ is given by Eq. \ref{eq:sim_eq}.

\begin{equation} \label{eq:sim_eq}
s = \frac{1}{2^{var{(P-Q)}}},
\end{equation}
where $var{(P-Q)}$ is the variance of the difference between $P$ and $Q$. Values of $s$ closer to 0 indicates low similarity and 1 high similarity. Similarity estimates using gyroscope grade estimations from different trips are averaged to get the similarity profile of a road segment. Regions of high similarity are selected as candidates based on a threshold ($s_{thresh}$), for the calculation of the offset. The optimal value of $s_{thresh}$ is empirically driven and set to 0.7 in the implementation of the system.

As illustrated in Fig. \ref{fig:similar}, road segment region from $\approx$ 350m to $\approx$ 850m has high similarity between estimations from Google elevation data and gyroscope. Regions within a road segment with similarity greater than $s_{thresh}$ are selected, and are denoted by $[x_{strt}, x_{end}]$, where $x_{strt}$ is starting point of the region and $x_{end}$ the end point.

To compute the offset, we pick samples of estimates of anchor snapshots that lie in the region of high similarity, i.e. samples that fall in $[x_{strt}, x_{end}]$. This vector of anchor snapshots is denoted by $A = \{a_{1}, a_{2}, a_{3}, \ldots, a_{n}\}$. The corresponding road grade estimations using Google elevation data are denoted by $G = \{g_{1}, g_{2}, g_{3}, \ldots, g_{n}\}$ are sampled. Finally, the offset in region of high similarity is calculated using Eq.~\ref{eq:offset}. 

\begin{equation} \label{eq:offset}
offset = mean(diff(A-G))
\end{equation}

Offsets calculated using Eq. \ref{eq:offset} are used to compensate for anchor snapshot estimations on a given road segment. We use the offset calculated in the nearest region of high similarity to estimate the corrected values of anchor snapshots. The result of applying the offset and correcting the drift is illustrated in Fig. \ref{fig:offsetted_out}. It is noteworthy that offset correction is valid for the entire course of the trip, as long as the orientation of smartphone does not change w.r.t the vehicle. 

\subsection{Aggregation Framework}\label{data_agg}
Due to its ubiquity and low-cost, smartphone as a sensing platform is an attractive option to develop crowd-sourced frameworks, to perform large scale sensing tasks. In our application, we can take advantage of observations of road grade from multiple sources to increase accuracy and robustness of our system. In a typical crowd-sourced system, sources can provide conflicting observations on an object due to varying QoI (Quality of Information) from multiple sources. A straight-forward way to handle the conflicts is to average the different observations on an object from multiple sources. However, simply averaging the observations ignore source reliability by considering them equal, and thus might suffer from incorporation of information from unreliable sources (with low QoI). Specific to our application, varying QoI of sources arises from factors such as varying a) quality of sensors in smartphones, b) suspension properties of different vehicles, c) inherent vibration of different vehicles, d) quality of phone holders used to mount the smartphone, etc. 

To handle the problem of varying QoI of sources, truth discovery methods \cite{li2014resolving,meng2015truth,li2016survey} are proposed, which take into account source reliability into data aggregation. Sources are assigned weights based on their QoI i.e. reliable sources with high QoI are given more weight and vice-versa. We leverage CRH~\cite{li2014resolving} to perform aggregation of observations from different sources. CRH formulates the observation conflicts from different sources as an optimization problem to minimize the overall weighted distance between the input and the estimated truths. 

As previously described, we design road grade estimation strategies using the accelerometer and gyroscope on commodity smartphones, exploiting the complementary nature of the two sensors. To recap, we opportunistically estimate grade snapshots using the accelerometer (anchor snapshots) when the vehicle dynamics are stable and correct the offsets in these estimates using information from Google elevation data, we also use the estimated anchor snapshots to compensate for the drift in estimation using the gyroscope to get the final road segment grade profile. We observe the following characteristics of estimations done using our proposed method:{\bf a) Sparse Observations from Accelerometer:} Since we opportunistically select anchor Snapshots during periods of stable vehicle dynamics, the grade estimations are sparse. The ``density'' and location of anchor snapshots on road segment is dependent on occurrence of stable driving events, thus is influenced by factors such as driving behavior of the user, traffic conditions, etc. For example, frequent rapid acceleration and deceleration events, due to the user being aggressive or traffic congestion while negotiating a road segment will result in fewer observations of anchor snapshots during a trip. Also, the beginning and end of a road segment generally has fewer observations because of rapid acceleration and deceleration (We divided the route into segments based on intersections). Fewer observations mean less information for drift correction to work with. {\bf b) Noisy estimations from Accelerometer and Gyroscope:}  Noise profile of accelerometer and gyroscope is dependent on factors such as quality of sensors on the phone, quality of phone holder, inherent vibrations of the vehicle, etc. For example, accelerometer and gyroscope estimations from a phone ``loosely'' mounted in the vehicle will experience more noise than that of a firmly mounted phone, especially on roads in poor condition.

We thus propose an aggregation framework. The aggregation of estimations from various trips is done in two steps a) {\em Aggregation of Anchor Snapshots}, and b) {\em Profile Aggregation}.

\subsubsection{Aggregation of Anchor Snapshots} \label{agg_snaps}

We aggregate estimations of anchor snapshots from various trips on a given road segment~(Fig.~\ref{fig:sys_arch}, top). The intuition is to handle the sparse estimations of anchor snapshots described above by increasing the density of anchor snapshots on a given road segment, using data from different trips. Furthermore, aggregation will compensate the effects of varying QoI from different sources. We divide the road segment into bins of length $bin_{acc}$ (set to 2m) and sample the observations from different trips in these bins. Observations in each bin are then aggregated to produce the final output. Fig. \ref{fig:agg_profiles} illustrates the aggregated result by applying CRH on the observations of anchor snapshots in each bin. Data from 15 trips is used to get the final aggregated result.

\subsubsection{Profile Aggregation}\label{agg_prof}

\begin{figure}
    \centering
    \begin{minipage}[t]{0.54\columnwidth}
        \centering
        \includegraphics[width=\textwidth]{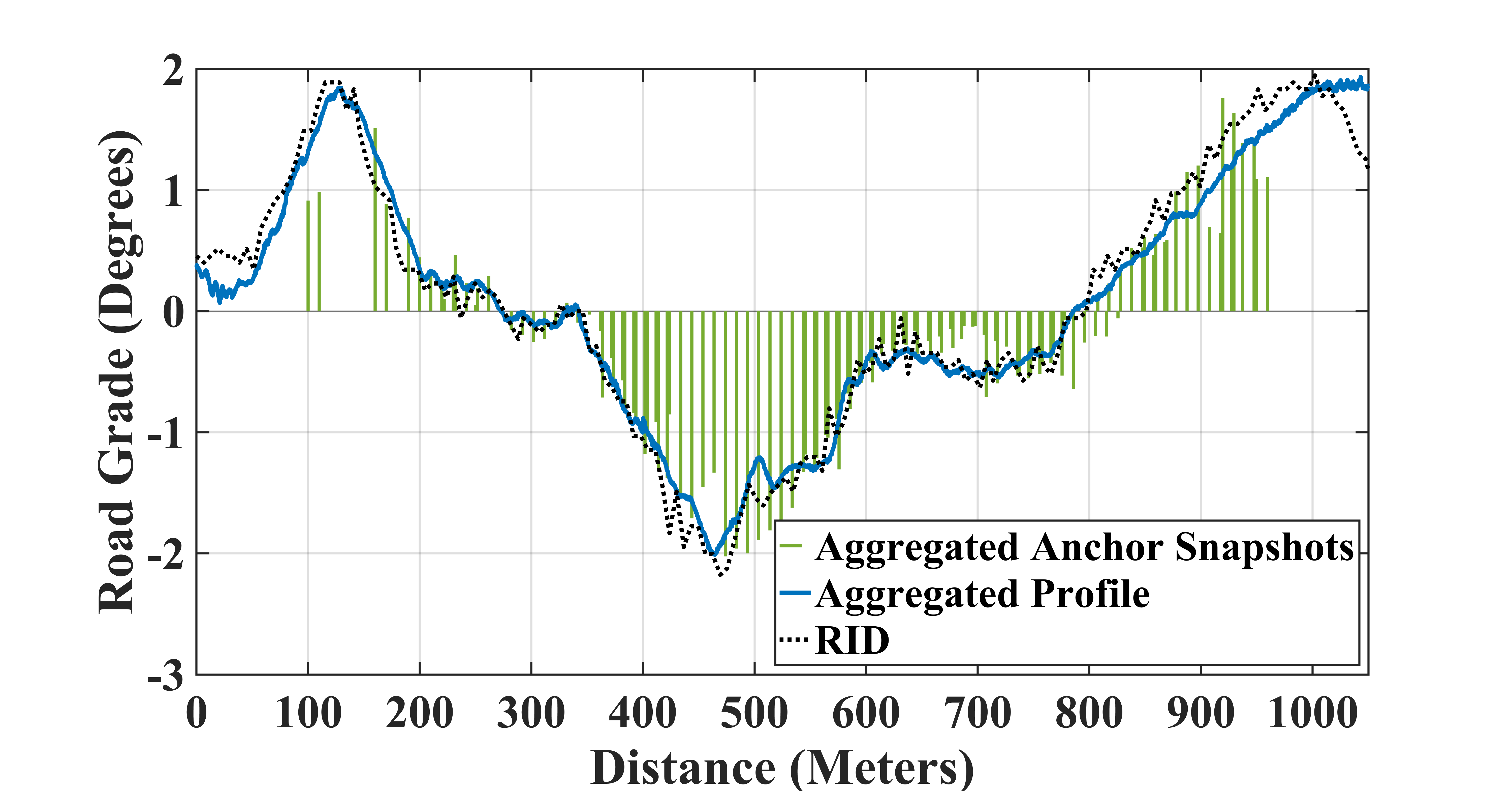} 
        \caption{Aggregation of anchor snapshots and individual profiles.}
        \label{fig:agg_profiles}
    \end{minipage}\hfill
    \begin{minipage}[t]{0.44\columnwidth}
        \centering
        \includegraphics[width=\textwidth]{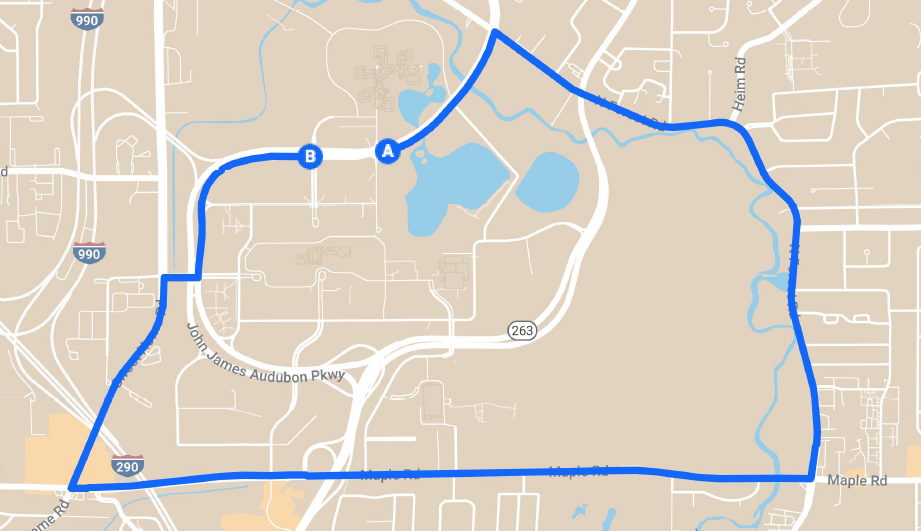} 
        \caption{Test route for evaluation.}
        \label{fig:test_route}
    \end{minipage}
\end{figure}

We apply the drift correction method described in Sec. \ref{drift_corr} on pitch estimations from gyroscope of different trips using the aggregated anchor snapshots on a given road segment~(Fig.~\ref{fig:sys_arch}, top). As done in Sec.~\ref{agg_snaps}, we divide the road segment into bins of length $bin_{corr-gyr}$. Compared to  $bin_{acc}$, $bin_{corr-gyr}$ is smaller (set to 20cm) due to presence of continuous observations from gyroscope. The corrected gyroscope estimations thus derived are finally aggregated into a single profile of road grade on a given road segment. The intuition is to ``average out'' the discrepancies in estimations using gyroscope due to factors such as quality of sensors on the phone, quality of phone holder, etc. Fig. \ref{fig:agg_profiles} illustrates the result of applying CRH on the observations in each bin.


\section{Evaluation}\label{eval}
\subsection{Experimental Setup}

\textbf{Data.} Data collection was done on a test route shown in Fig. \ref{fig:test_route} using 3 different smartphones: Nexus 5, Nexus5x and Google Pixel XL. We leverage VehSense, an android application, that was developed for data collection. Smartphones were fixed in arbitrary orientation in the vehicle using phone-holders at various locations such as wind-shield, air-conditioning vents, etc. Specifically we collect the following time-series samples: a) 3-axis angular velocity data from the gyroscope, b) 3-axis acceleration data from the accelerometer, c) GPS data including latitude, longitude, speed and bearing, d) 3-axis magnetic field data from the magnetometer, and e) vehicle speed data from the OBD-II scanner, paired with the smartphone via bluetooth.

Data collection was done over the course of one month in May 2019 in diverse traffic scenarios, ranging from high traffic during peak hours ($\approx$ 9am-10am and $\approx$ 5pm-6pm) to low traffic ($\approx$ 3pm-4pm and $\approx$ 10pm - 11pm). The collected data is comprised of 15 trips distributed over 5 participants. The vehicles used for data collection were Kia Optima Hybrid - 2012 (2400cc), Subaru Forester - 2016 (2500cc), Toyota Camry - 2011 (2500cc), Toyota Corolla - 2014~(1800cc) and Nissan Altima - 2005 (3500cc). For coordinate alignment, each participant was asked to keep the vehicle stationary for $\approx$ 30 seconds in the parking lot, before starting the trip at Point A on the map~(Fig.~\ref{fig:test_route}). The participants were asked to drive naturally on the route, which is $\approx$ 9 km. The test route includes a variety of road types such as freeway, streets, secondary highway, as well as including road structures such as bridge, underpass, etc.

\textbf{Comparison.} We evaluate the performance of our proposed method by comparing it to the approaches as listed below.

\begin{itemize}
	\item \textbf{3RD}~\cite{yang2016low}: As per our knowledge, 3RD is the only smartphone-based solution to estimate road grade. It relies on fusion of estimates from magnetometer, accelerometer, and gyroscope. The intuition behind 3RD's fusion framework is: a) estimate from magnetometer and accelerometer is accurate in the long run; b) estimate from gyroscope is precise in short intervals; and c) combine low-pass filtered magnetometer and accelerometer estimates with high-pass filtered gyroscope estimates to produce the final road grade estimations.

	\item \textbf{Google Elevation}: We leverage Google Elevation API \cite{google} to sample elevation on the test route. To do this, a) we perform map-matching~\cite{newson2009hidden} on GPS traces from the smartphones using Google Map matching API~\cite{google_snap} to compensate for noise in the GPS data. Map-matching snaps the GPS coordinates from the smartphone to the center-line of the nearest road segment; b) the output of the map-matched coordinates are queried against the Google elevation API to get the elevation data for the route, and c) finally, the grade profile of the route is generated using Eq. \ref{eq:grade_google}, where $E_{D}$ is the elevation gain over distance $D$.
\end{itemize}

\textbf{Groundtruth.}  For groundtruth, we use data from the Road Inventory Database~\cite{rid}, which includes information on road geometry features (curvature, grade and superelevation) of $\sim$25,000 directional miles of roadway in six sites in USA. The data is collected using ARAN (Automatic Road Analyzer)\cite{aran}, a specialized instrumented vehicle with high-grade IMU's, laser scanners, high-precision GPS, and camera.

\textbf{Evaluation Measure}: We use the following to compare the performance of our system with the baselines:

\begin{itemize}
\item {\bf Absolute Error (AE):} It is the absolute value of difference between our estimation and the ground-truth at a point on the road.

\item {\bf Gradient Error (GE):} It is the absolute value of difference of change of road grade per unit distance between our estimation and the ground-truth. This metric will indicate the performance of various methodologies in capturing the shape of the road profile.
\end{itemize}

We list the abbreviations used in the paper:

\begin{itemize}
    \item  {\bf Indiv-Anch:} Anchor snapshots estimated using method in Sec.~\ref{snaps} and corrected by applying offsets using method in Sec.~\ref{goog_elev}.
    \item {\bf Indiv-Prof:} Grade Profiles generated using Indiv-Anch by applying drift correction using method in Sec.~\ref{drift_corr}~(i.e. without applying data aggregation method in Sec.~\ref{data_agg}).
    \item {\bf Indiv-Prof-GPS:} Indiv-Prof using velocity from smartphone's GPS instead of OBD-II.
    \item {\bf Agg-Anch:} Aggregated anchor snapshots generated by aggregating Indiv-Anch using method in Sec.~\ref{agg_snaps}.
    \item {\bf Agg-Anch-Prof:} Grade profiles generated using Agg-Anch by applying drift correction using method in Sec.~\ref{drift_corr}.
    \item {\bf Agg-Prof-Final:} Aggregated grade profiles generated by aggregating Agg-Anch-Prof using method in Sec.~\ref{agg_prof}.
    \item {\bf Agg-Prof-Final-GPS:} Agg-Prof-Final using velocity from smartphone's GPS instead of OBD-II.
    \item {\bf Indiv-Anch-No-Off:} Anchor snapshots estimated using method in Sec.~\ref{snaps}, but without applying offsets from Sec. \ref{goog_elev}.
    \item {\bf Agg-Anch-No-Off:} Aggregated anchor snapshots generated by aggregating Indiv-Anch-No-Off using method in Sec.~\ref{agg_snaps}.
    \item {\bf Agg-Anch-Prof-No-Off:} Grade profiles generated using Agg-Anch-No-Off by applying drift correction using method in Sec.~\ref{drift_corr}.
    \item {\bf Agg-Prof-Final-No-Off:} Aggregated grade profiles generated by aggregating Agg-Anch-Prof-No-Off using method in Sec.~\ref{agg_prof}.
    \end{itemize}



\subsection{Overall Performance}
\begin{figure*}
	\centering
	\begin{subfigure} [b] {0.33\textwidth}
		\includegraphics[width=\textwidth]{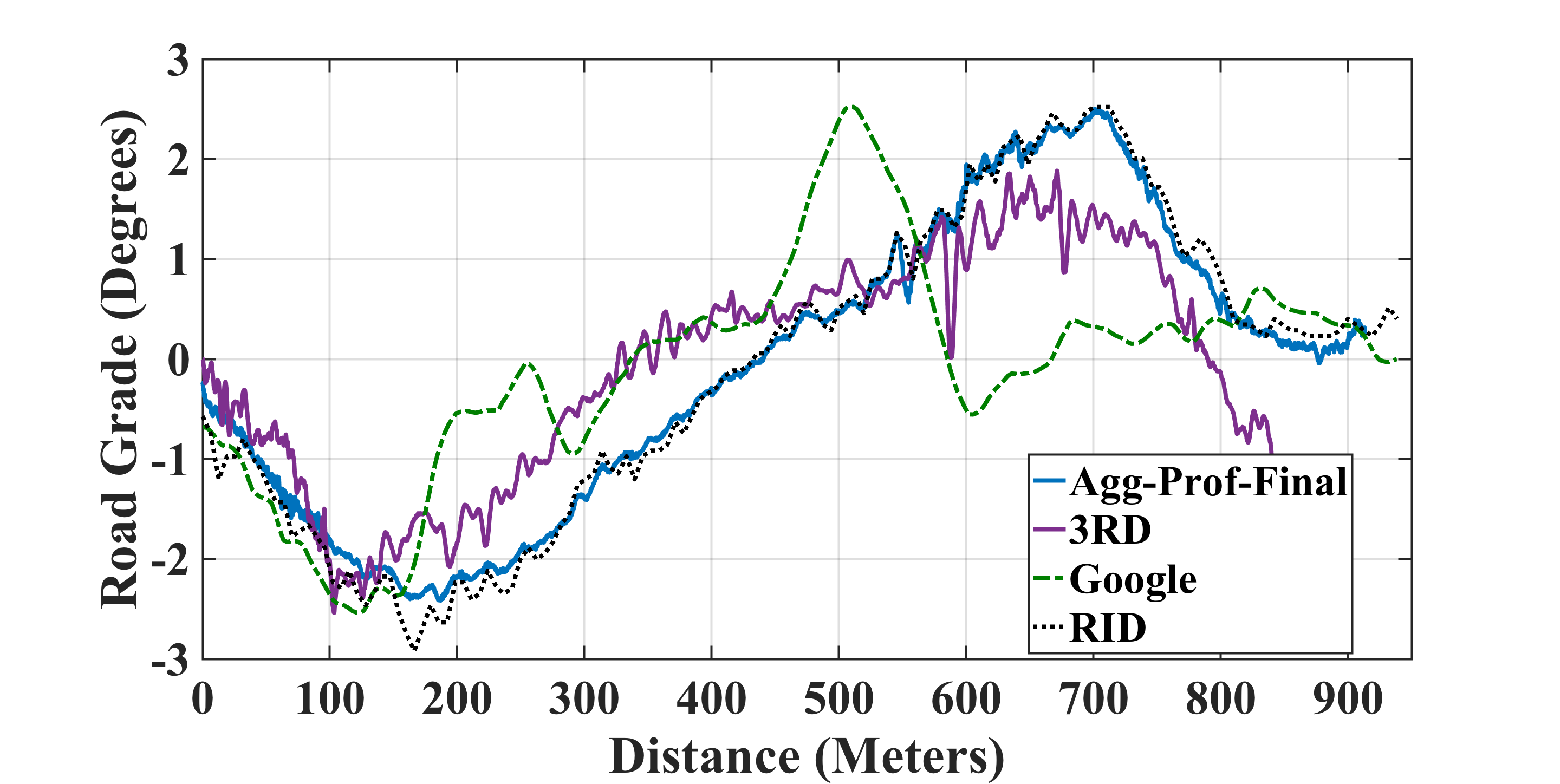}
		\caption{Road Segment 4.}
		\label{fig:comp_maple}
	\end{subfigure}
	\begin{subfigure}[b] {0.33\textwidth}
		\includegraphics[width=\textwidth]{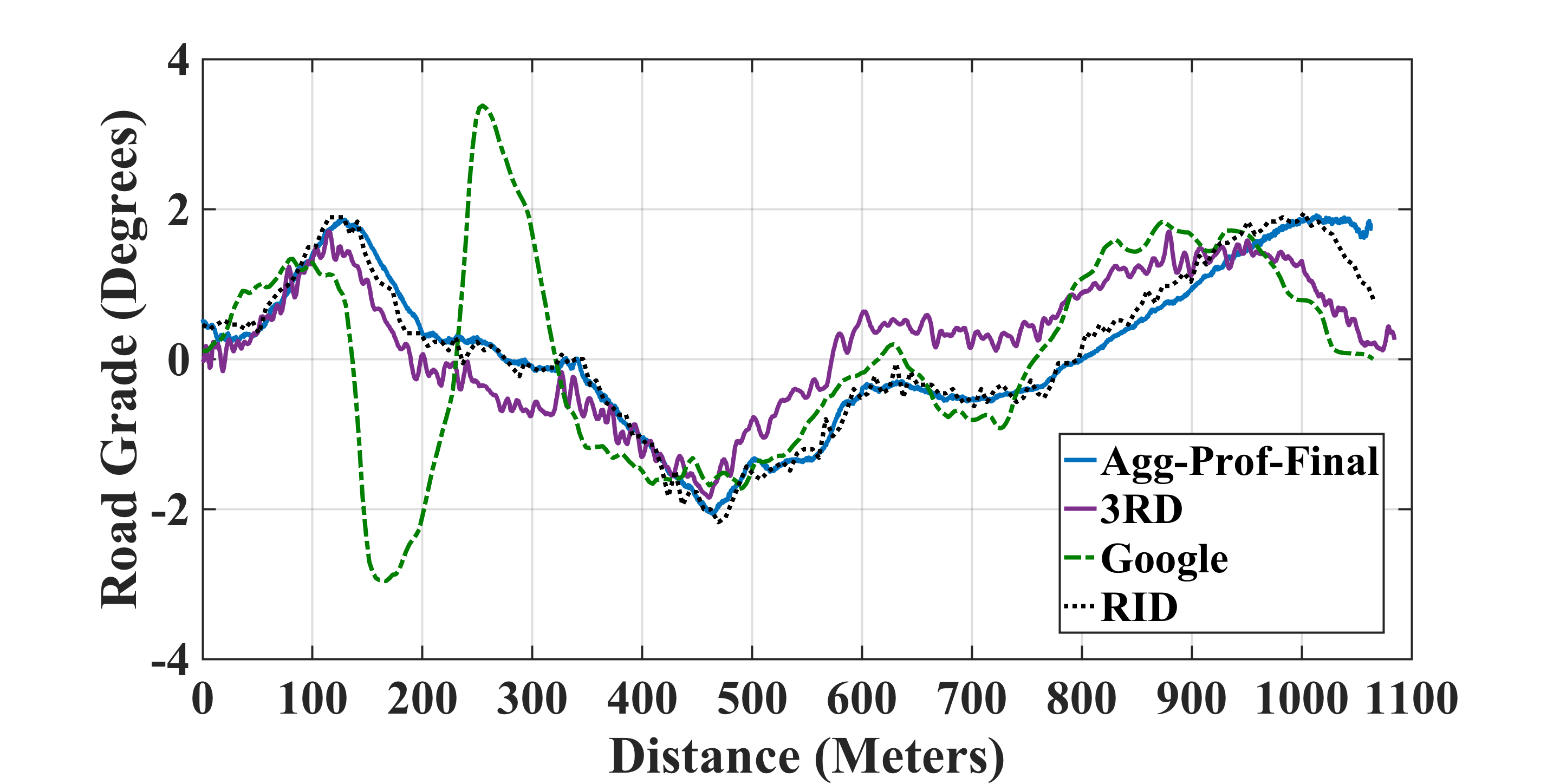}
		\caption{Road Segment 5.}
		\label{fig:comp_6664}
	\end{subfigure}
	\begin{subfigure}[b] {0.33\textwidth}
		\includegraphics[width=\textwidth]{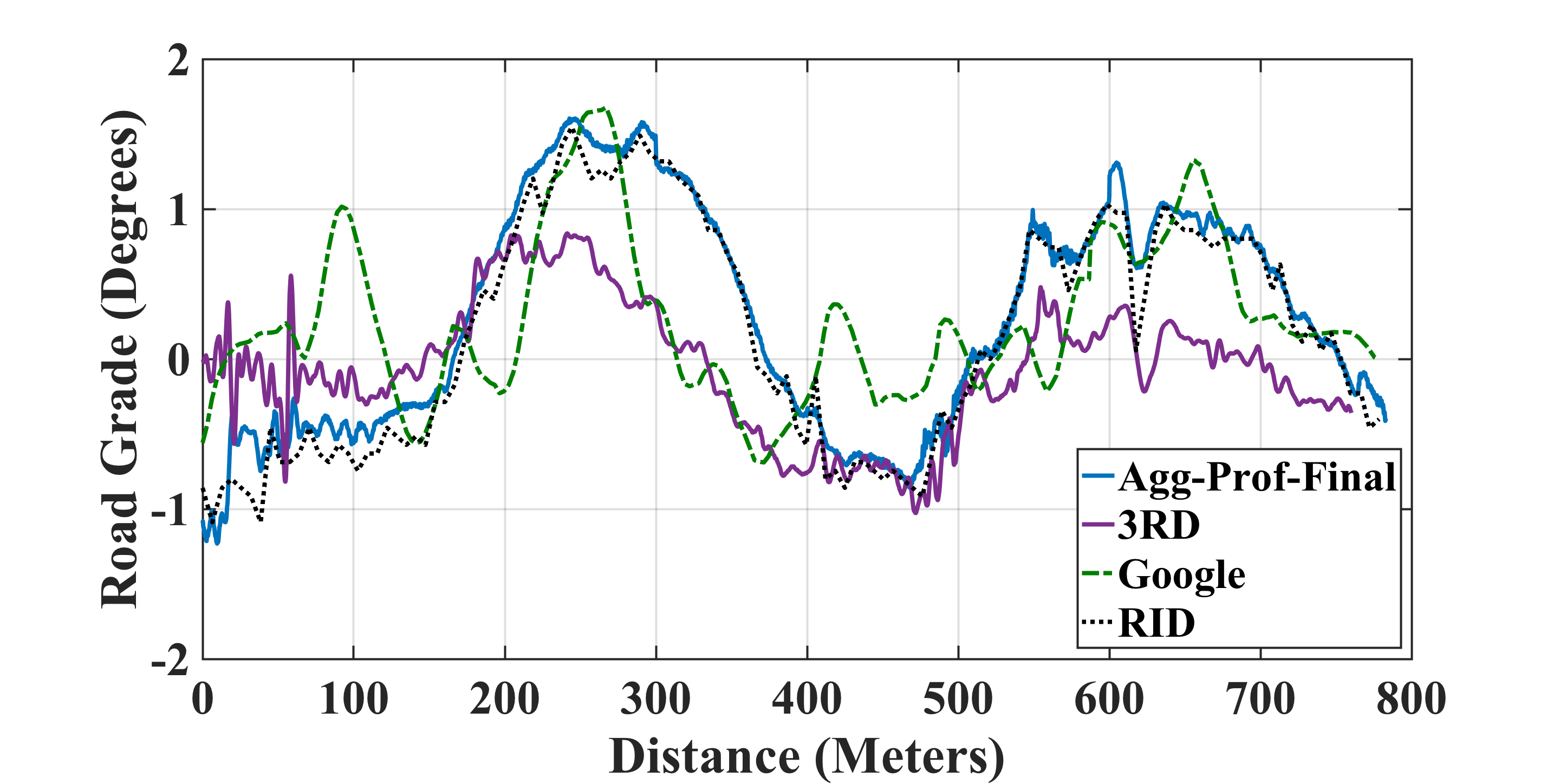}
		\caption{Road Segment 1.}
		\label{fig:comp_8850}
	\end{subfigure}
	\caption{Comparison of road grade profiles estimated using our approach with baselines on different road segments.}
	\label{fig:comparison_profiles}
\end{figure*}

In this section, we analyze the overall performance of our system and compare it with the baselines. Fig. \ref{fig:comp_maple}, \ref{fig:comp_6664}, and  \ref{fig:comp_8850} show the comparison of road grade profile estimated using our approach and the baselines on different road segments. The low resolution accuracy of Google elevation data is evident from the figures, thus establishing the motivation to use ground sensing techniques in the first place. Specifically, the performance of Google elevation data suffers on road segments that do not follow the profile of surrounding terrain or are occluded due to surrounding infrastructure. For example, Road Segment 4 (as shown in Fig.~\ref{fig:comp_maple}) is a bridge, and Road Segment 5 (as shown in Fig. \ref{fig:comp_6664}) is an underpass. Due to the low resolution, google elevation misses data points on the road. Thus, the error magnifies on an elevated structure such as a bridge, where some elevation points lie on terrain below the bridge, resulting in an inaccurate estimation of grade from $\approx$ 150m - $\approx$850m. Furthermore, susceptibility to occlusions of remote sensing techniques is illustrated in  Fig.~\ref{fig:comp_6664}. The peaks at $\approx$ 200m and $\approx$ 300m are result of an elevated road that goes over the profiled road segment.

\begin{figure}
	\centering
	\begin{subfigure} [b] {0.38\textwidth}
		\includegraphics[width=\textwidth]{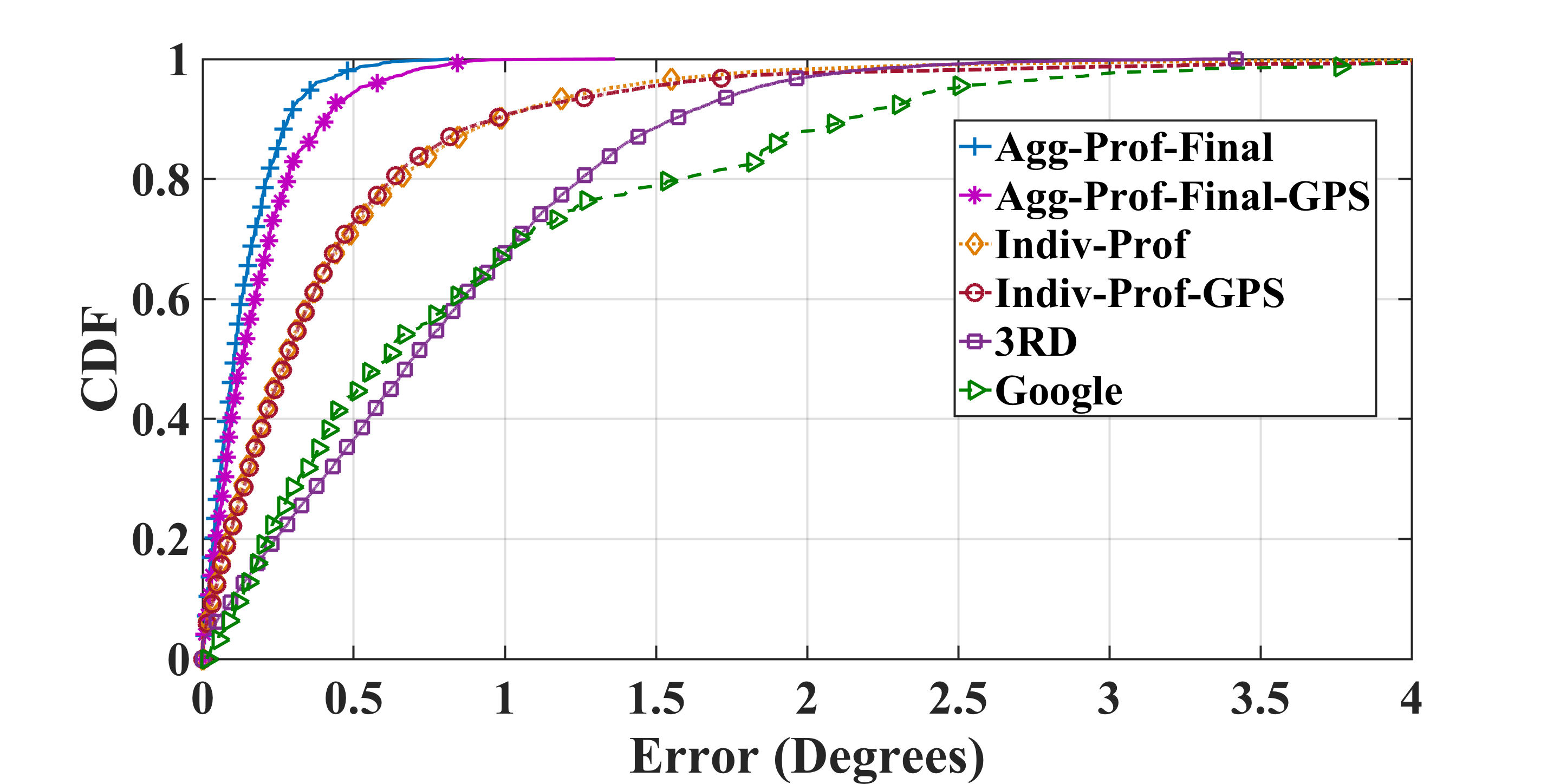}
		\caption{Distribution of Absolute Error (AE).}
		\label{fig:over_all_ae_dist}
		
	\end{subfigure}
	\begin{subfigure}[b] {0.38\textwidth}
		\includegraphics[width=\textwidth]{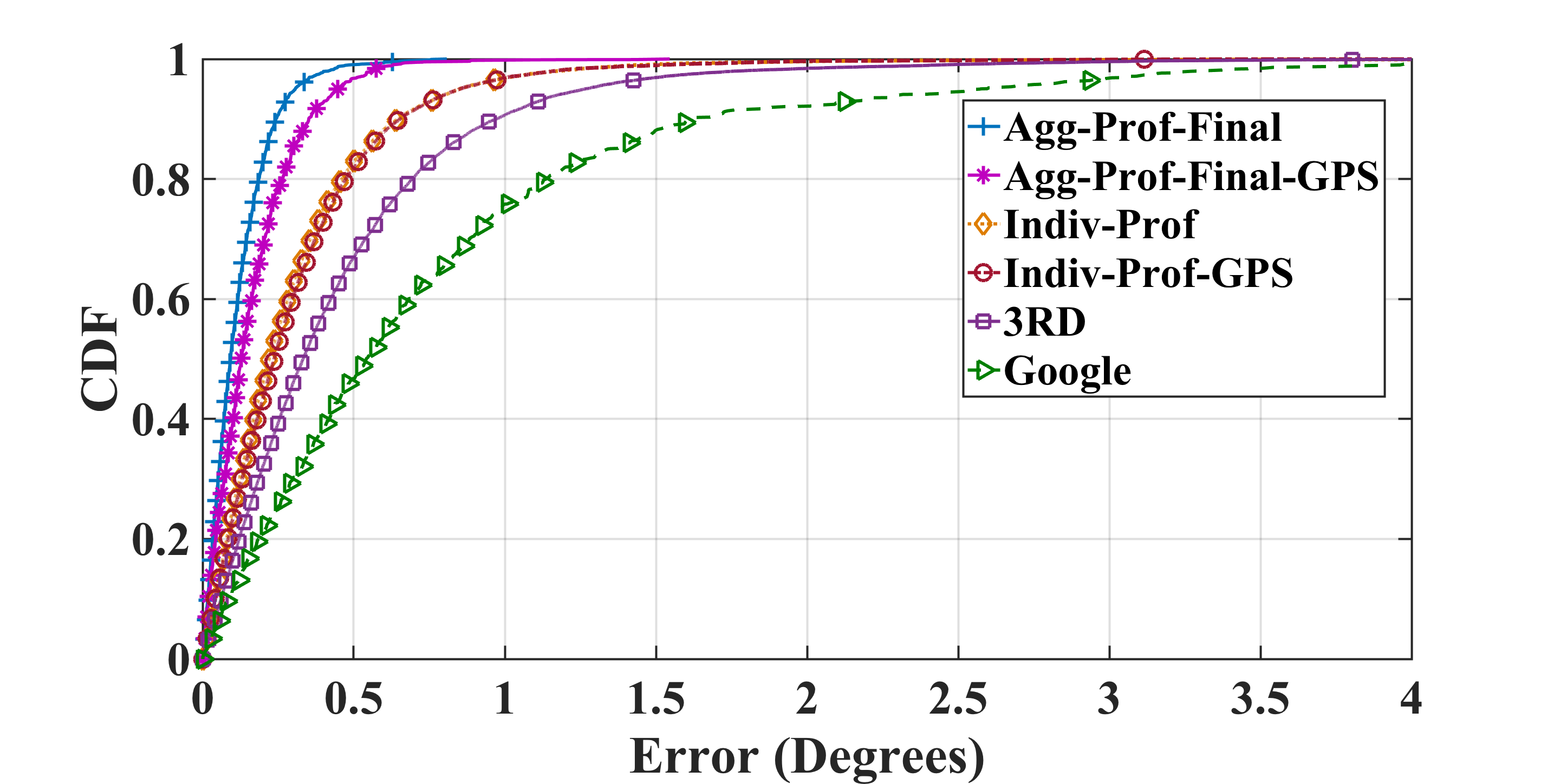}
		\caption{Distribution of Gradient Error (GE)}
		\label{fig:over_all_ge_dist}
	\end{subfigure}
	\caption{Comparison of error distributions of different methods.}
	\label{fig:over_all_comparison_cdf}
\end{figure}

Fig .\ref{fig:over_all_ae_dist} shows the statistical comparison of absolute error (AE) of different approaches on the entire test route. Both 3RD and our approach outperform grade estimation using Google elevation data, demonstrating accuracy gains of ground sensing using high resolution data over a remote sensing technique. The 50\% and 90\% absolute error (AE) of road grade estimated using Google elevation data are 0.6$^\circ$ and 2.3$^\circ$, respectively. 3RD's 50\% and 90\% AE are 0.67$^\circ$ and 1.57$^\circ$, respectively. In comparison, our approach without aggregation (labeled ``Indiv-Prof") has 50\% and 90\% AE of 0.27$^\circ$ and 0.97$^\circ$, respectively. 

Gains in accuracy of 3RD and our approach over grade estimation using Google elevation data are also observed in Fig~ .\ref{fig:over_all_ge_dist}, which illustrates gradient error (GE) of different approaches on the entire test route. The 50\% and 90\% GE of road grade estimated using Google elevation data are 0.54$^\circ$ and 1.6$^\circ$, respectively. 3RD's 50\% and 90\% AE are 0.33$^\circ$ and 0.97$^\circ$, respectively. In comparison, our approach without aggregation (labeled ``Indiv-Prof") has 50\% and 90\% AE of 0.22$^\circ$ and 0.64$^\circ$, respectively. As compared to AE, gains in accuracy of 3RD and our approach are more pronounced for GE. As mentioned in Sec.~\ref{goog_elev} and analyzed in Sec.~\ref{goog_elev_anal}, point estimations of road grade using Google elevation data can be reasonably accurate in certain regions of road segment. However, due to the low resolution of Google elevation data, the over-all shape accuracy of grade estimations is poor.

Our approach without applying data aggregation (labeled ``Indiv-Prof") outperforms 3RD in both Absolute Error and Gradient Error, demonstrating the impact of intelligently combining the estimations from accelerometer and gyroscope based on understanding of their error characteristics in a driving environment. In particular, 3RD suffers in accuracy due to : \textbf{a) Susceptibility to vehicle dynamics}: 3RD's grade estimation framework is based on fusing estimates from accelerometer, magnetometer, and gyroscope. To exploit the complimentary nature of sensors, 3RD adds/fuses the low-frequency component of magnetometer and accelerometer estimation with high-frequency component of gyroscope's estimate. However, optimal parameter setting (e.g., cut-off frequency for the filters) is difficult to achieve because of varying road types and dynamic noise profiles of smartphone's sensors. This makes 3RD susceptible to both absolute and gradient errors, especially due to pollution of estimation from accelerometer during periods when the dynamics of vehicle are not stable. Our approach, on the other hand, relies primarily on estimation from gyroscope which is not prone to errors induced by vehicle dynamics.\textbf{b) Dynamic Gyroscope Drift}: 3RD assumes that the drift remains constant during the trip. Estimation of the drift is done whenever the vehicle is stationary and the subsequent gyroscope samples are compensated using the estimation. However, as discussed in Sec.~\ref{drift_corr}, in a dynamic driving environment drifts can vary over time. Thus, 3RD fails to capture the dynamic nature of the drift. Our approach, on the other hand, benefits from ``on-the-go'' drift correction.
	
Next we analyse the impact of data aggregation. Gains in accuracy due to aggregation of estimations from various trips is evident from the results as shown in Fig~.\ref{fig:over_all_ae_dist} and \ref{fig:over_all_ge_dist}. The 50\% and 90\% AE of aggregated road grade estimations are 0.1$^\circ$ and 0.28$^\circ$, respectively. The 50\% and 90\% GE are 0.09$^\circ$ and 0.24$^\circ$, respectively. The results indicate the power of data aggregation, which essentially ``averages out'' errors in estimation due to varied data quality of smartphone sensors. In particular, varied data quality is due to factors such as inherent vibrations of the vehicle, varied suspension properties of the vehicle, quality of phone-holder, quality of sensors on smartphone, etc.

Finally, Fig.~\ref{fig:over_all_ae_dist} and \ref{fig:over_all_ge_dist} demonstrate the performance of the proposed system using velocity from smartphone's GPS (labeled ``Agg-Prof-Final-GPS'' and ``Indiv-Prof-GPS'') instead of OBD-II. The 50\% and 90\% AE of aggregated road grade estimations are 0.13$^\circ$ and 0.41$^\circ$, respectively. The 50\% and 90\% GE are 0.13$^\circ$ and 0.35$^\circ$, respectively. The results indicate that desirable performance can be achieved by the proposed system using smartphone as a standalone sensing platform, without relying on data from any external sensors. The marginal drop in accuracy is mainly attributed to occasional in-accurate velocity from GPS due to poor signal.

\subsection{Impact of ${acc_{thresh}}$ and ${jerk_{thresh}}$}
We also analyze the response of our system to various settings of ${acc_{thresh}}$ and ${jerk_{thresh}}$, which are used to filter periods when the dynamics of the vehicle are stable to estimate anchor snapshots. Lower thresholds place tighter bound on the dynamics of the vehicle and vice-versa. Table. \ref{fig:anch_acc_density} illustrates the impact of various settings of ${acc_{thresh}}$ and ${jerk_{thresh}}$ on the accuracy and density of anchor snapshots. It can be seen that lower thresholds result in better accuracy. 
However, strict thresholds result in lower density of anchor snapshots, thus demonstrating trade-off between accuracy and quantity of information. 


Next, we analyze the impact of ${acc_{thresh}}$ and ${jerk_{thresh}}$ on the aggregated estimations of road grade. Since the performance of ``On the go Drift Correction'' mechanism is dependent on both the accuracy and density of anchor snapshots, strict settings of ${acc_{thresh}}$ and ${jerk_{thresh}}$ does not guarantee better performance. The settings ${acc_{thresh}} = 0.7$, ${jerk_{thresh}} = 0.15$ (Mean AE = 0.13$^\circ$) and ${acc_{thresh}} = 0.9$, ${jerk_{thresh}} = 0.2$ (Mean AE = 0.14$^\circ$) result in better accuracy than the setting ${acc_{thresh}} = 0.5$, ${jerk_{thresh}} = 0.1$ (Mean AE = 0.16$^\circ$).

\begin{table}[]\resizebox{0.3\textwidth}{!}{
	\begin{tabular}{|c|l|l|l|l|l|}
		\hline
		\begin{tabular}[c]{@{}c@{}}$acc_{thresh}$(m/sec$^2$), \\ $jerk_{thresh}$(m/sec$^3$)\end{tabular}         & \begin{tabular}[c]{@{}l@{}}0.5, \\ 0.1\end{tabular} & \begin{tabular}[c]{@{}l@{}}0.7, \\ 0.15\end{tabular} & \begin{tabular}[c]{@{}l@{}}0.9, \\ 0.2\end{tabular} & \begin{tabular}[c]{@{}l@{}}1.3, \\ 0.3\end{tabular} & \begin{tabular}[c]{@{}l@{}}2.0, \\ 0.5\end{tabular} \\ \hline
		\begin{tabular}[c]{@{}c@{}}Density(Number \\ of Snapshots/500m)\end{tabular} & 11.47                                               & 13.26                                                & 16.76                                               & 18.89                                               & 20.18                                               \\ \hline
		\begin{tabular}[c]{@{}c@{}}Mean AE of \\ Indiv-Anch\end{tabular}       & 0.26$^\circ$                                                  & 0.29$^\circ$                                                   & 0.31$^\circ$                                                  & 0.33$^\circ$                                                  & 0.36$^\circ$                                                  \\ \hline
		\begin{tabular}[c]{@{}c@{}}Mean AE of \\ Agg-Prof-Final\end{tabular} & 0.16$^\circ$                                                  & 0.13$^\circ$                                                   & 0.14$^\circ$                                                  & 0.16$^\circ$                                                  & 0.17$^\circ$                                                  \\ \hline
	\end{tabular}}
\caption{Impact of $acc_{thresh}$ and $jerk_{thresh}$ on accuracy and density of anchor snapshots.}
\label{fig:anch_acc_density}
\end{table}

\subsection{Offset Correction Using Google Elevation Data}\label{goog_elev_anal}

\begin{figure}
	\centering
	\begin{subfigure} [b] {0.33\textwidth}
		\includegraphics[width=\textwidth]{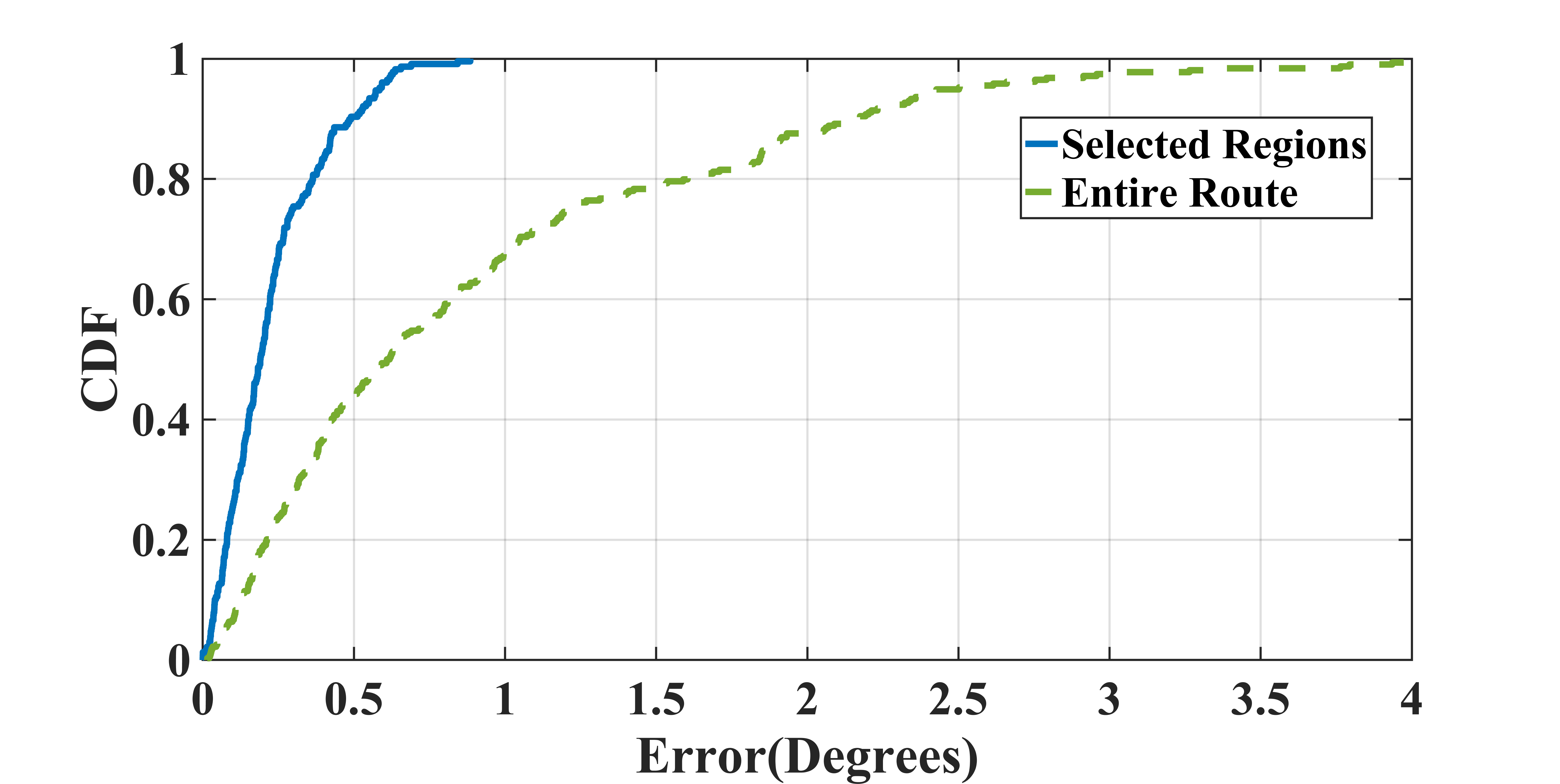}
		\caption{Comparison of Absolute Error (AE) distributions of estimated grade using Google elevation data in selected regions and entire route.}
		\label{fig:goog_selected}
	\end{subfigure}
	\begin{subfigure}[b] {0.33\textwidth}
		\includegraphics[width=\textwidth]{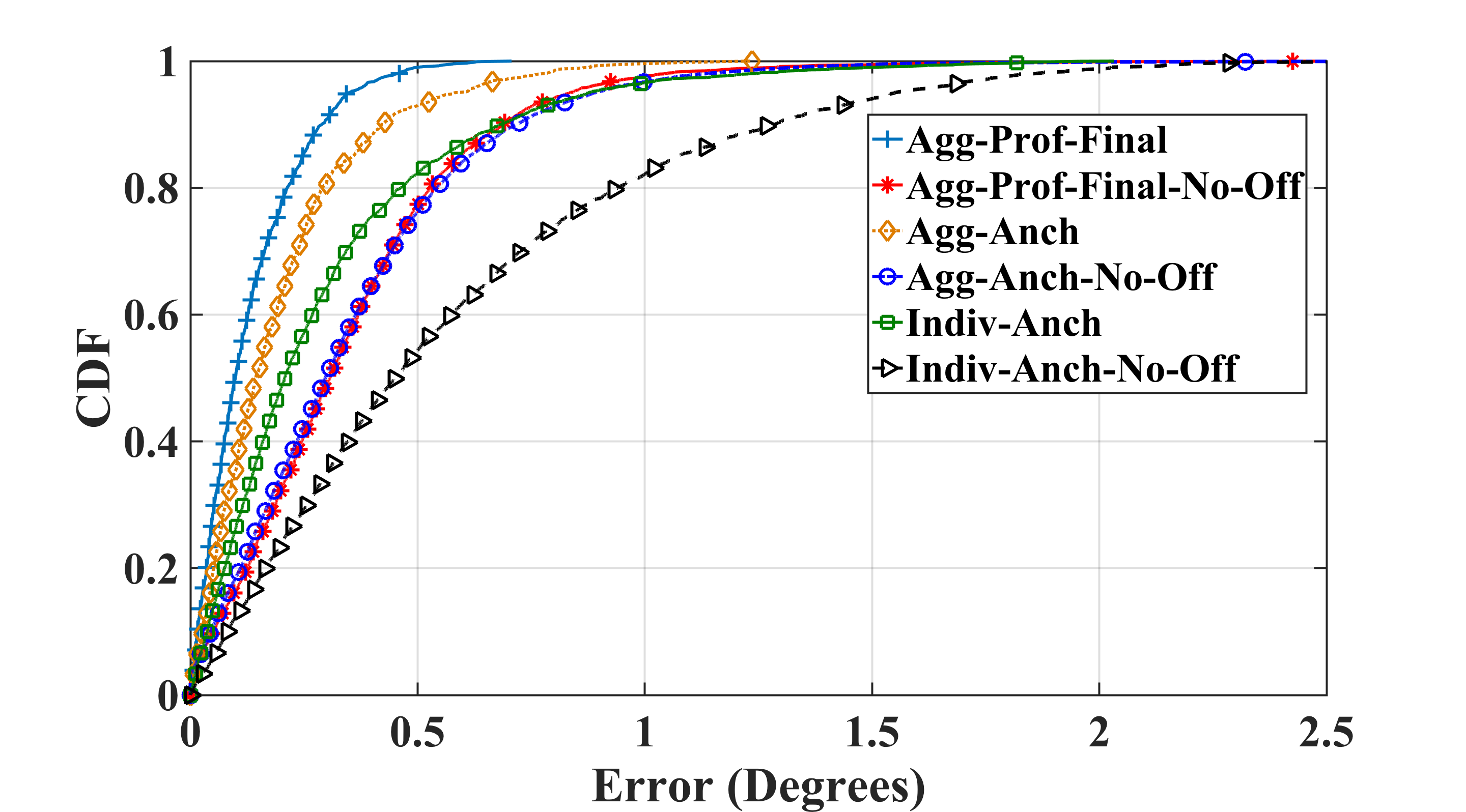}
		\caption{Impact on Absolute Error (AE) of offset correction methodology.}
		\label{fig:off_corr_anal}
	\end{subfigure}
	\caption{Evaluating impact of offset correction using Google elevation data.}
	\label{fig:offset_anal}
\end{figure}

Fig. \ref{fig:goog_selected} illustrates the efficacy of methodology described in Sec. \ref{goog_elev} in extracting regions of high accuracy grade estimation using Google elevation data. 
The results indicate that in selected regions, accuracy of Google elevation data is impressive and can act as a reliable input to correct the offsets using the methodology described in Sec. \ref{goog_elev}. The total length of extracted regions of high accuracy is $\approx$ 1200m, which is $\approx$ 14\% of the length of the entire test route.

Fig. \ref{fig:off_corr_anal} illustrates the impact of offset correction on accuracy of the system. In general, offset correction results in accuracy gains. Applying offsets on anchor snapshots improves the 50\% and 90\% AE by 0.24$^\circ$ and 0.59$^\circ$~(labeled ``Indiv-Anch-No-Off'' and `Indiv-Anch''), respectively. 50\% and 90\% AE gains for aggregated anchor snapshots are 0.15$^\circ$ and 0.28$^\circ$~(labeled ``Agg-Anch-No-Off'' and ``Agg-Anch''). 50\% and 90\% AE gains for aggregated grade profiles are 0.21$^\circ$ and 0.4$^\circ$~(labeled ``Agg-Prof-Final-No-Off'' and ``Agg-Prof-Final'').


\begin{figure*}
	\centering
	
	\begin{subfigure} [b] {0.33\textwidth}
		\includegraphics[width=\textwidth]{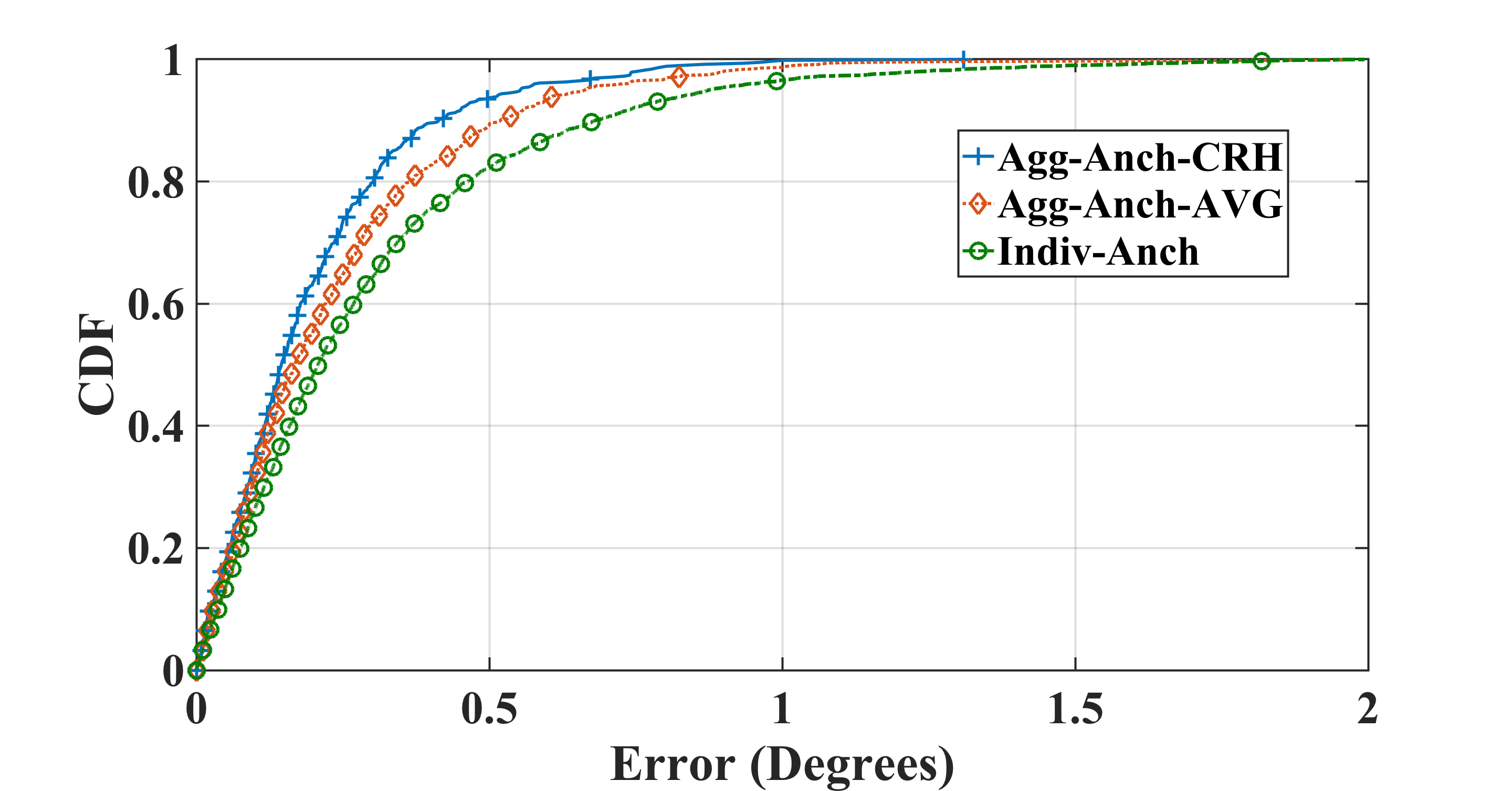}
		\caption{Comparison of Absolute error (AE) distributions of aggregated anchor snapshots.}
		\label{fig:agg_anch_snaps_anal}
	\end{subfigure}
	\begin{subfigure}[b] {0.33\textwidth}
		\includegraphics[width=\textwidth]{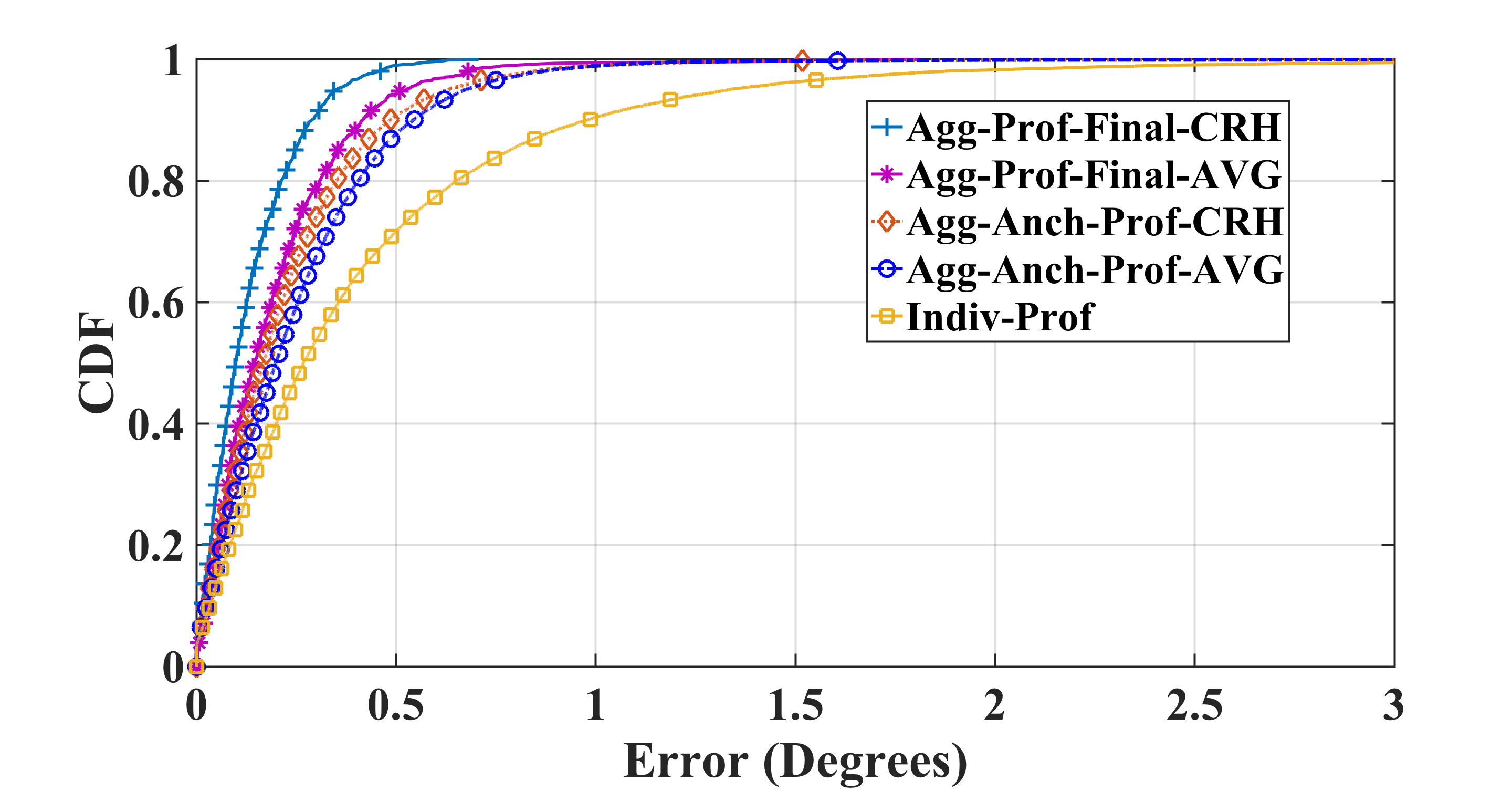}
		\caption{Comparison of Absolute Error (AE) distributions of aggregated grade profiles.}
		\label{fig:agg_final_anal}
	\end{subfigure}
	\begin{subfigure}[b] {0.33\textwidth}
		\includegraphics[width=\textwidth]{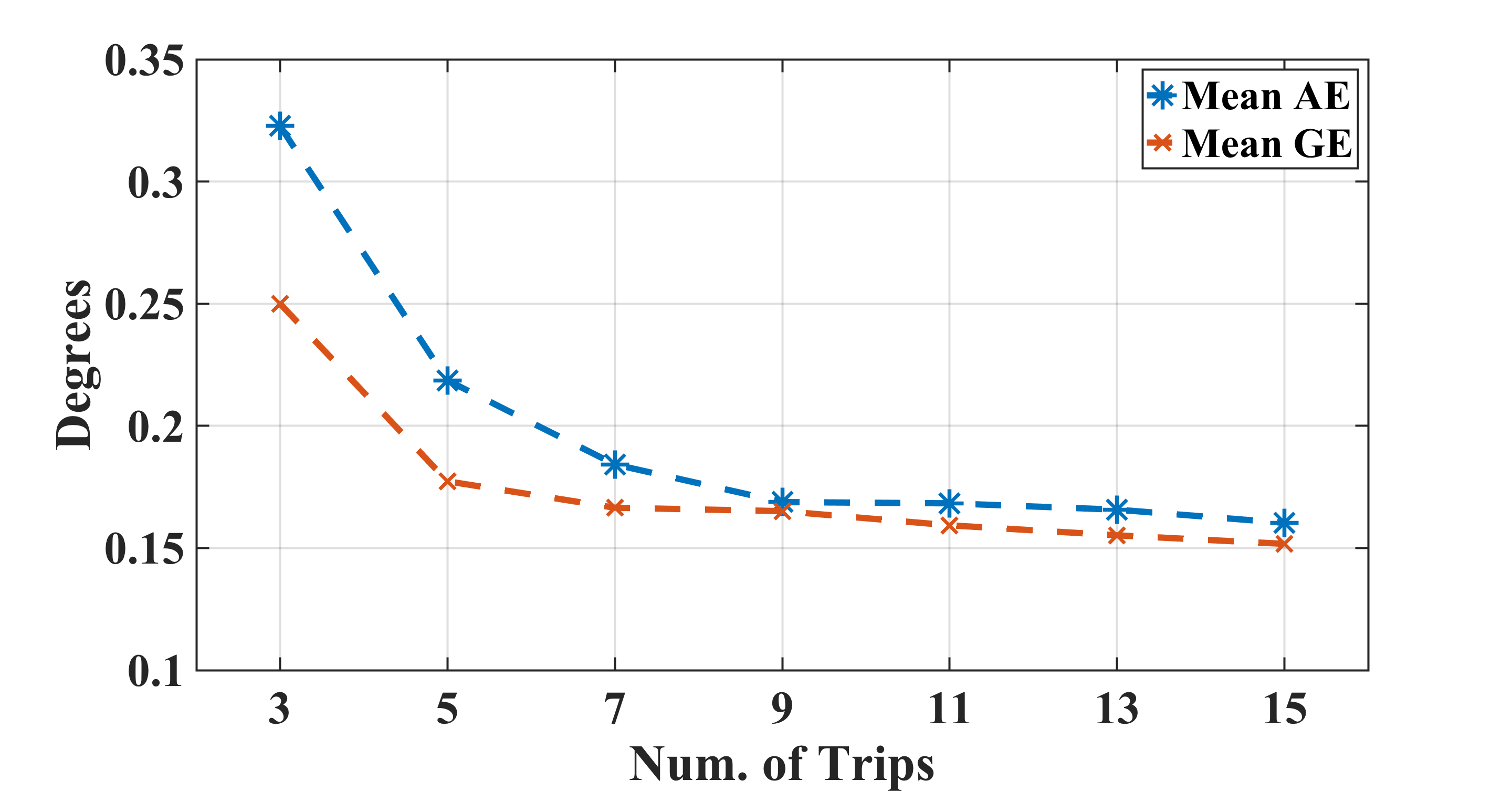}
		\caption{Performance of aggregation framework with varying amount of data.}
		\label{fig:num_trips}
	\end{subfigure}
	
	\caption{Evaluation of the data aggregation framework.}
	\label{fig:aggregation_analysis}
\end{figure*}

\subsection{Data Aggregation}

Fig. \ref{fig:agg_anch_snaps_anal} illustrates the impact of performing aggregation on estimates of anchor snapshots. In general, aggregated estimations of anchor snapshots result in accuracy gains. The 50\% and 90\% gain in accuracy of aggregated anchor snapshots (labeled ``Agg-Anch-CRH'') over ``Indiv-Anch'' is 0.06$^\circ$ and 0.31$^\circ$, respectively. 
Furthermore, the performance of aggregated estimation computed using CRH is better than simply averaging the anchor snapshots (labeled ``Agg-Anch-AVG``). 

Gains in accuracy due to aggregation can also be seen in Fig.~\ref{fig:agg_final_anal}, which shows the AE distributions for the estimated road grade profiles. Furthermore, CRH consistently performs better than average as it takes into account source reliability by assigning more weight to reliable sources. The reason for better performance of aggregated estimations is two-fold. Firstly, aggregation of estimations compensates the varied quality of data from different sources. In particular, the varied quality of data is due to {a)} heterogenous smartphones/sensors, {b)} varied driving behaviour, and {c)} varied noise profile of the smartphone sensor due to quality of phone-holder, inherent vibration of the vehicle, etc. Secondly, aggregation results in an increased density of anchor snapshots, which results in more information for ``On-the-go-drift correction'' (Sec.~\ref{drift_corr}) module to work with.

What's more, Fig. \ref{fig:num_trips} demonstrates trend in accuracy of the data aggregation framework w.r.t. data used in terms of the number of trips. With more data, the accuracy of the system improves, as expected.

\section{Discussion and Future Work }\label{discuss}

We discuss some of the limitations of the proposed approach: a) The proposed method works with the assumption that the orientation of the phone does not change w.r.t vehicle during estimation of road grade. If orientation change occurs, the system will have to wait for the next stationary and acceleration phase on a straight road to perform coordinate alignment (Sec.~\ref{coor_align}), to be ready to estimate road grade again. Although, it is possible to extract sensor traces on a particular road segment with no significant orientation changes ~\cite{mohan2008nericell} from a pool of participants in a crowd-sourcing system (e.g., while using navigation services people prefer to keep their phones secured using phone holders), the assumption limits the scope of applicability of the proposed system. We would like to investigate the above ideas and the impact of phone-usage/movements on our system in our future work. b) We make use of Google elevation data to correct the offsets in estimation of grade by extracting regions of high accuracy (Sec.~\ref{goog_elev}). However, it might be possible that in certain areas (e.g., mountain roads, where most of the roads do not follow the surrounding terrain and severity of occlusions is more pronounced), regions of high accuracy does not exist.

For future work, we would like to explore feasibility of using generated road grade profiles in developing techniques for improved localization and navigation.

\section{Conclusion}\label{conclude}
This paper presents a novel, cost-efficient and easily deployable crowd-sourcing system for road grade estimation using Smartphones. Deriving insights from analysis of smartphone sensor's error characteristics in a dynamic driving environment, we intelligently integrate data from accelerometer, gyroscope and Google elevation to estimate road grade. Finally, we crowd-source observations from various sources to improve accuracy and robustness of the system. The experiment results demonstrate that the proposed method provides considerable improvement over existing solutions.

\begin{acks}
This work was supported in part by the US National Science Foundation under Grants CNS-1652503 and CNS-1737590. The views and conclusions contained in this document are those of the authors and should not be interpreted as representing the official policies, either expressed or implied, of NSF.
\end{acks}

\bibliographystyle{ACM-Reference-Format}
\bibliography{ref}

\end{document}